\documentclass[11pt,twoside]{article}
\usepackage{amsmath}
\usepackage{amssymb}
\usepackage{amscd}
\usepackage{mathrsfs}
\usepackage{graphicx}
\usepackage{epstopdf}
\usepackage[usenames]{color}
\usepackage{wrapfig}
\usepackage{boxedminipage}
\renewcommand{\a}{\alpha}
\renewcommand{\b}{\beta}
\newcommand{\cart}[1]{\operatorname*{\times}_{\!{\scriptscriptstyle #1}} }
\newcommand{\cd}{\mathfrak{d}}
\newcommand{\cev}[1]{\smash{\overset{\smash{{}_{\gets}}}{#1}}}
\newcommand{\codiv}{\nabla\!{\cdot}}
\newcommand{\comp}{\mathbin{\raisebox{1pt}{$\scriptstyle\circ$}}}
\newcommand{\C}{{\boldsymbol{C}}}
\newcommand{\Ccal}{{\mathcal{C}}}
\newcommand{\Cs}{{\rlap{\lower3pt\hbox{\textnormal{\LARGE \char'040}}}{\Gamma}}{}}
\renewcommand{\d}{\delta}
\newcommand{\dbk}{\displaybreak[2]\\}
\newcommand{\dde}[2]{\frac{\partial #1}{\partial #2}}
\newcommand{\de}{\partial}
\newcommand{\dx}{\dO\xx}
\newcommand{\dy}{\dO\yy}
\newcommand{\dz}{\dO\zz}
\newcommand{\dH}{\mathrm{d}_{\sst{\mathrm H}}}
\newcommand{\dO}{\mathrm{d}}
\newcommand{\D}{{\boldsymbol{D}}}
\newcommand{\DF}{\mathfrak{D}\!\F}
\newcommand{\DO}{\mathrm{D}}
\newcommand{\E}{{\boldsymbol{E}}}
\newcommand{\Ecal}{{\mathcal{E}}}
\newcommand{\End}{\operatorname{End}}
\newcommand{\F}{{\boldsymbol{F}}}
\newcommand{\fnb}[1]{[\![#1]\!]}
\newcommand{\g}{\gamma}
\newcommand{\G}{\Gamma}
\newcommand{\ih}{\tfrac{\iO}{2}}
\newcommand{\iO}{\mathrm{i}}
\newcommand{\Id}[1]{{1\!\!1}\!{}_{#1}{}}
\newcommand{\Ical}{{\mathcal{I}}}
\newcommand{\iI}[2]{{}_{#1}^{\phantom{#1}\!#2}}
\newcommand{\iIi}[3]{{}_{#1\phantom{#2}\!\!#3}^{\phantom{#1}\!#2}}
\newcommand{\Ii}[2]{{}^{#1}_{\phantom{#1}\!#2}}
\newcommand{\IiI}[3]{{}^{#1\phantom{#2}\!\!#3}_{\phantom{#1}\!#2}}

\newcommand{\jO}{\mathrm{j}}
\newcommand{\JO}{\mathrm{J}}
\newcommand{\Jcal}{{\mathcal{J}}}
\newcommand{\kk}{{\mathsf{k}}}
\renewcommand{\k}{\kappa}
\newcommand{\K}{{\boldsymbol{K}}}
\newcommand{\lfr}{\mathfrak{l}}
\newcommand{\Lcal}{{\mathcal{L}}}
\newcommand{\Lie}{\mathfrak{L}}
\newcommand{\LO}{\mathrm{L}}
\newcommand{\mdots}{{\cdot}{\cdot}{\cdot}}
\newcommand{\M}{{\boldsymbol{M}}}
\newcommand{\nasl}{{\rlap{\raise1pt\hbox{\,/}}\nabla}}
\newcommand{\oh}{\tfrac{1}{2}}
\newcommand{\oq}{\tfrac{1}{4}}
\newcommand{\onto}{\rightarrowtail}
\newcommand{\ost}[1]{\overset{{}_{{\,}_*}}{#1}}
\renewcommand{\OE}[1]{\Omega^{#1}\!\E{}}
\newcommand{\pint}{{\scriptscriptstyle\mathord{\rfloor}}}
\newcommand{\Pcal}{{\mathcal{P}}}
\newcommand{\qRq}{{\quad\Rightarrow\quad}}
\newcommand{\rdg}{{\textstyle\sqrt{{\scriptstyle|}g{\scriptstyle|}}\,}}
\newcommand{\rrdg}{{\scriptstyle\sqrt{{\sst|}g{\sst|}}}}
\newcommand{\RR}{{\mathbb{R}}}
\newcommand{\sref}[1]{\S\ref{#1}}
\newcommand{\sst}{\scriptscriptstyle}
\newcommand{\sA}{{\scriptscriptstyle A}}
\newcommand{\sB}{{\scriptscriptstyle B}}
\newcommand{\sI}{{\scriptscriptstyle I}}
\newcommand{\spec}[1]{{}_{\sst{\mathrm{#1}}}}
\newcommand{\sGa}{{\scriptstyle\Gamma}}
\newcommand{\Scal}{{\mathcal{S}}}
\newcommand{\ten}[1]{\operatorname*{\otimes}_{\!{\scriptscriptstyle #1}} }
\newcommand{\tn}{{\,\otimes\,}}
\newcommand{\Tcal}{{\mathcal{T}}}
\newcommand{\TO}{\mathrm{T}}
\newcommand{\TS}{\TO^{*}\!}
\newcommand{\ul}{\underline}
\newcommand{\up}{{\scriptscriptstyle\uparrow}}
\newcommand{\Ucal}{{\mathcal{U}}}
\newcommand{\VO}{\mathrm{V}}
\newcommand{\VE}{{\VO\!\E}}
\newcommand{\VF}{\VO\!\F}
\newcommand{\we}{{\,\wedge\,}}
\newcommand{\weu}[1]{{\wedge^{\!#1}}}
\newcommand{\W}{{\boldsymbol{W}}}
\newcommand{\xx}{{\mathsf{x}}}
\newcommand{\yy}{{\mathsf{y}}}
\newcommand{\Y}{{\boldsymbol{Y}}}
\newcommand{\zz}{{\mathsf{z}}}
\newtheorem{proposition}{Proposition}[section]
\newtheorem{theorem}{Theorem}[section]
\newcommand{\remark}{\smallbreak\noindent{\bf Remark.}}
\title{Covariant-differential formulation of Lagrangian field theory}
\date{{\small v4: April 24, 2018} }
\author{{D.\ Canarutto} \\[6pt]
{\small\it Dipartimento di Matematica e Informatica ``U.~Dini'', }\\
{\small\it Via S. Marta 3, 50139 Firenze, Italia}\\
{\small email:~daniel.canarutto@unifi.it}\\
{\small http://www.dma.unifi.it/\char126 canarutto}}
\begin{document}
\maketitle \thispagestyle{empty}
\begin{abstract}\noindent
Building on the Utiyama principle
we formulate an approach to Lagrangian field theory
in which exterior covariant differentials of vector-valued forms
replace partial derivatives, in the sense that they take up the role
played by the latter in the usual jet bundle formulation.
Actually a natural Lagrangian can be written as a density
on a suitable ``covariant prolongation bundle'';
the related momenta turn out to be natural vector-valued forms,
and the field equations can be expressed in terms
of covariant exterior differentials of the momenta.
Currents and energy-tensors naturally also fit into this formalism.
The examples of bosonic fields and spin one-half fields,
interacting with non-Abelian gauge fields, are worked out.
The ``metric-affine'' description of the gravitational field
is naturally included, too.
\end{abstract}

\bigbreak\noindent

\smallbreak\noindent
\thispagestyle{empty}
\tableofcontents

\section*{Introduction}

Let \hbox{$\E\onto\M$} be a vector bundle.
The notion of \emph{[exterior] covariant differential} of $\E$-valued
exterior forms on $\M$,
defined by means of a linear connection of $\E$,
can be viewed as a generalization of the standard covariant derivative;
it has been variously present in the literature for several years,
possibly in relation to ideas by Koszul~\cite{Kos50,Kos60}.
More recently, the realization that the Fr\"olicher-Nijenhuis bracket
yields a natural framework for dealing with such notions
has suggested further generalizations and a systematic study by various
authors~\cite{FroNij56,FroNij60,Nij72,MaMo83b,ManMod84,Mod91,
Michor01,KolarMichorSlovak93,Grabowski13}.
In particular, it has been observed that the curvature tensor
of any connection can be regarded as the covariant differential
of the connection with respect to itself.

Thus, in consideration of the
\emph{Utiyama principle}~\cite{Utiyama56,KolarMichorSlovak93,Janyska07},
covariant differentials may provide a convenient setting
for natural gauge field theories.
Actually one finds, in the General Relativistic literature,
a formulation of Lagrangian field theory on a gravitational background
in which covariant derivatives with respect to the spacetime connection
replace partial derivatives~\cite{LandauLifchitz68,HE73}.
While that formulation explicitely considers only fields with spacetime indices,
it is not difficult to become convinced that,
even when the fields have further ``internal degrees of freedom'',
labeled by fiber indices that are not ``soldered'' to spacetime,
such limitation does not invalidate the essential results
regarding the properties of the stress-energy tensor
of matter fields (the right-hand side of the Einstein equation).

The main purpose of this paper is to present a generalization
of that approach,
and to explore the extent up to which the role of partial derivatives
can be taken up by covariant differentials.
The field of an essential gauge field theory is to be described
as a couple consisting of a \emph{matter field} and a \emph{gauge field},
namely \hbox{$(\phi,\k):\M\to\F\equiv\E\cart{\M}\C$}
where \hbox{$\C\onto\M$} is a bundle of linear connections of $\E$.
Denoting by $\dO_\k$ the covariant differential with respect to $\k$\,,
we get a ``covariant prolongation''
$$(\dO_\k\phi,\dO_\k\k):\M\to(\TS\M\tn\E)
\times_{\!\sst\M}(\weu2\TS\M\tn\End\E)$$
that is suitable for taking up the role of the fields' jet prolongations
used in the standard formulation of Lagrangian field theory.
Indeed, the Lagrangian density can be expressed as a function
on a ``covariant prolongation space'' $\DF$
(rather than a function on the first jet prolongation space $\JO\F$),
and we show that various fundamental constructions and results
of the standard theory have a counterpart in this new formulation.

In particular, momenta turn out to be geometrically
well-defined objects on $\DF$.
Infinitesimal variations can be introduced as morphisms on $\DF$,
and the field equations can be expressed in terms of covariant differentials
of the momenta evalued through the covariant prolongations
of the fields (theorem~\ref{theorem:fieldequations}).
Thus, the field equations can be explicitely written in coordinate-free form.
Currents and canonical energy-tensors can also be revisited
in terms of objects defined on $\DF$ .

Above, the base $\M$ can be a generic manifold.
Then (\sref{s:Field theory in spacetime})
we consider the case when $\M$ is a Lorentzian spacetime|%
whose structure is regarded as a fixed gravitational background|%
and the matter field can have spacetime or spinor indices.
These behave differently from the other internal degrees of freedom,
but the formalism can be naturally adapted to this situation
and yields analogous results.
In particular the field equations can be written again
in terms of covariant differentials of the momenta.
Moreover a \emph{generalized replacement theorem} holds,
so that the covariant differentials can be replaced
by covariant divergences (up to torsion terms).

The usual argument regarding the stress-energy tensor is then reviewed
in hopefully clearer terms from a geometric point of view.
Finally we consider concrete examples,
giving the explicit coordinate-free expressions
and the coordinate expressions
of the momenta, the field equations and the energy-tensors
for a field of arbitrary integer spin
and for a field of spin one-half,
both interacting with non-Abelian gauge fields.
We also show how the ``metric-affine'' treatment of the gravitational field
fits into the covariant-differential formulation.

\section{Covariant differential}\label{s:Covariant differential}
\subsection{Fr\"olicher-Nijenhuis bracket and covariant differential}
\label{ss:Froelicher-Nijenhuis bracket and covariant differential}

Let $\M$ be a sufficiently regular real manifold.
A \emph{tangent-valued (t.v.) $r$-form} on $\M$ is a section
\hbox{$\M\to\weu{r}\TS\M\tn\TO\M$}.
If $\Phi$ is a t.v.\ $r$-form and $\Psi$ is a t.v.\ $s$-form then
their \emph{Fr\"olicher-Nijenhuis bracket} \hbox{$\fnb{\Phi,\Psi}$}
is a t.v.\ \hbox{$(r\,{+}\,s)$}-form%
~\cite{MaMo83b,ManMod84,Michor01,KolarMichorSlovak93}.

Considering the FN-bracket on a fibered manifold \hbox{${\mathsf{p}}:\E\onto\M$},
we are mostly interested in ``basic'' t.v.\ forms
\hbox{$\E\to\weu{r}\TS\M\ten{\E}\TO\E$}.
In particular, a connection of \hbox{$\E\onto\M$}
can be regarded as a special, basic t.v.\ $1$-form
\hbox{$\k:\E\to\TS\M\ten{\E}\TO\E$}.
The \emph{covariant differential}~\cite{MaMo83b,ManMod84,Mod91}
associated with $\k$ acts
on a t.v.\ $r$-form $\Phi$ as \hbox{$\dO_\k\Phi\equiv\fnb{\k,\Phi}$}\,,
which turns out to be a basic t.v.\ $(r\,{+}\,1)$-form
if $\Phi$ is basic.

In this paper we will mostly deal with the case
when \hbox{$\E\onto\M$} is a vector bundle.
We will explicitly indicate the base manifold in a fiber tensor product
only when it is not $\M$.
We will use the shorthand
$$\OE{r}\equiv\weu{r}\TS\M\tn\E~,$$
and note that
$$\E\cart{\M}\OE{r}\cong\weu{r}\TS\M\ten{\E}\VE\onto\E$$
can be regarded as the bundle
of vertical-valued basic $r$-forms on $\E$
(as now the vertical subspace \hbox{$\VE\subset\TO\E$}
can be identified with \hbox{$\E\cart{\M}\E$}).
In particular, a section \hbox{$\phi:\M\to\E$} yields the section
$$\check\phi:\E\to\OE{0}\equiv\VE:y\mapsto(y,\phi({\mathsf{p}}(y))$$
and \hbox{$\dO_\k\check\phi$} is essentially $\nabla\!_\k\phi$\,,
namely the covariant differential can be regarded as a natural generalization
of the usual covariant derivative.

The latter statement can be intended in a broader sense by observing
that the curvature tensor associated with $\k$ is the section
$$\rho\equiv-\dO_\k\k\equiv-\fnb{\k,\k}~,$$
which in turn can be regarded~\cite{C16c} as the covariant derivative of $\k$
with respect to a certain ``overconnection'',
i.e.\ a connection (associated with $\k$ itself)
of the bundle of linear connections of $\E$.

Let $\bigl(\xx^a\,,\,\yy^i\bigr)$ be linear fibered coordinates on $\E$.
We denote the induced fiber coordinates on \hbox{${\otimes}^r\TS\M$}
and \hbox{${\otimes}^r\TS\M\tn\E$} by the shorthands
$$\zz_{a_1\dots a_r}\equiv\de\xx_{a_1}\tn\mdots\tn\de\xx_{a_r}~,\qquad
\zz\iI{a_1\dots a_r}i\equiv\zz_{a_1\dots a_r}\tn\yy^i~,$$
and use the same symbols for their restrictions to $\weu{r}\TS\M$ and $\OE{r}$.
Moreover we set
$$\dx_{a_1\dots a_r}\equiv \zz_{a_1\dots a_r}|\dO^m\xx \quad
\text{(interior product)},$$
where \hbox{$\dO^m\xx\equiv\dx^1\we\mdots\we\dx^m$}, \hbox{$m\equiv\dim\M$}.
We obtain a handy ``complementary notation'' 
for basic forms of degree $m\,{-}\,r$\,,
namely if \hbox{$\xi:\E\to\OE{m-r}$} then we write
\begin{align*}
&\xi=\xi^{a_1\dots a_r\,i}\,\dx_{a_1\dots a_r}\tn\de\yy_i=
\xi\iI{a_{r+1}\dots a_m}i\,\dx^{a_{r+1}}\we\mdots\we\dx^{a_m}\tn\de\yy^i~,
\\[6pt]
&\xi^{a_1\dots a_r\,i}=
\tfrac{1}{r!}\,\varepsilon^{a_1\dots a_r\,a_{r+1}\dots a_m}\,\xi\iI{a_{r+1}\dots a_m}i~,
\qquad
\xi\iI{a_{r+1}\dots a_m}i=
\tfrac1{(m-r)!}\,\varepsilon_{a_1\dots a_ra_{r+1}\dots a_m}\,\xi^{a_1\dots a_r\,i}~.
\end{align*}

It is not difficult to show that
$$\dx^b\we\dx_{a_1\dots a_r}=
\tfrac1{(r-1)!}\,\d^b_{[a_r}\,\dx_{a_1\dots a_{r-1}]}~,$$
where the above anti-symmetrization over the indices \hbox{$a_1\dots a_r$}
does \emph{not} include normalizing factorials.
Using this identity one computes the ``complementary notation''
coordinate expression of $\dO_\k\xi$ for \hbox{$\xi:\E\to\OE{m-r}$},
obtaining
$$\dO_\k\xi=
r\,\bigl(\de_{a_r}\xi^{a_1\dots a_{r-1}a_r\,i}
+\de_j\xi^{a_1\dots a_{r-1}a_r\,i}\,\k_{a_r}^j
-\xi^{a_1\dots a_{r-1}a_r\,j}\,\de_j\k_{a_r}^i\bigr)\,
\dx^{\phantom{a}}_{a_1\dots a_{r-1}}\tn\de\yy_i~.$$

\subsection{Generalized replacement principle}
\label{ss:Generalized replacement principle}

In the sequel we will often deal with the special situation when $\k$
is a linear connection and \hbox{$\xi:\M\to\OE{m-r}$}
(the condition that the ``source'' manifold be $\M$
typically arises when $\xi$ is obtained
by evaluating some object defined on $\E$ through a field).
Then we get the simplified expression
$$\dO_\k\xi=
r\,\bigl(\de_{a_r}\xi^{a_1\dots a_{r-1}a_r\,i}
-\xi^{a_1\dots a_{r-1}a_r\,j}\,\k\iIi{a_r}ij\bigr)\,
\dx^{\phantom{a}}_{a_1\dots a_{r-1}}\tn\de\yy_i~.$$

Let now $\G$ be a linear connection of \hbox{$\TO\M\onto\M$}.
The covariant derivative $\nabla\xi$ with respect to the couple $(\G,\k)$|%
see~\sref{ss:Gauge field theory on a General Relativistic background}|is
a section
$$\M\to\TS\M\ten{\M}\weu{m-r}\TS\M\tn\E~.$$
Under \emph{antisymmetrization}, this yields the ``covariant divergence''
$$\codiv\xi=r\,\nabla\!_{a_r}\xi^{a_1\dots a_{r-1}a_r\,i}\,
\dx_{a_1\dots a_{r-1}}\tn\de\yy_i\;:\;
\M\to\weu{m-r+1}\TS\M\tn\E\equiv\OE{m-r+1}~,$$
which we want to compare with $\dO_\k\xi$\,.

The relation between $\codiv\xi$ and $\dO_\k\xi$
turns out to be governed by the torsion $T$ of $\G$,
with components \hbox{$T\Ii a{bc}\equiv\G\iIi cab-\G\iIi bac$}\,.
Consider the sections
$$\tau\we\xi\,,\;T\,{\bar\wedge}\,\xi\;:\;\M\to\OE{m-r+1}~,$$
where $\tau$ is the \emph{torsion $1$-form},
with components \hbox{$\tau_a=T\Ii b{ab}$}\,, and the symbol ${\bar\wedge}$
stands for exterior product \emph{followed} by interior product.
We get the coordinate expressions
\begin{align*}
\tau\we\xi&=r\,\xi^{a_1\dots a_r\,i}\,\tau_{a_r}\,\dx_{a_1\dots a_{r-1}}\tn\de\yy_i~,
\\[6pt]
T\,{\bar\wedge}\,\xi&=
r\,(r\,{-}\,1)\,T\Ii{a_{r-1}}{bc}\,\xi^{a_1\dots a_{r-2}\,bc\,i}\,
\dx_{a_1\dots a_{r-1}}\tn\de\yy_i~.
\end{align*}

A coordinate computation then yields:
\begin{proposition}\label{prop:genrepprinc}
({\bf\emph{generalized replacement principle}}).
We have
$$\codiv\xi=\dO_\k\xi-\tau\we\xi-\oh\,T\,{\bar\wedge}\,\xi~.$$
\end{proposition}

Thus we see that $\codiv\xi$ depends on $\G$ only via its torsion,
and coincides with $\dO_\k\xi$ if \hbox{$T=0$}\,.
In particular:
\smallbreak\noindent$\bullet$%
~if \hbox{$r=0$}\,, \hbox{$\xi=\xi^i\,\dO^m\xx\tn\de\yy_i$}\,,
then \hbox{$\codiv\xi=0$}\,;

\smallbreak\noindent$\bullet$%
~if \hbox{$r=1$}\,,
\hbox{$\xi=\xi^{a\,i}\,\dx_a\tn\de\yy_i$}\,, then
\hbox{$\codiv\xi=(\de_a\xi^{ai}\,{-}\,\k\iIi aij\,\xi^{aj}
\,{-}\,\tau_a\,\xi^{ai})\,\dO^4\xx\tn\de\yy_i=\dO_\k\xi-\tau\we\xi$}\,;

\smallbreak\noindent$\bullet$%
~if \hbox{$r=2$}\,, 
\hbox{$\xi=\xi^{ab\,i}\,\dx_{ab}\tn\de\yy_i$}\,, then
$$\codiv\xi=2\,\bigl(\de_a\xi^{ba\,i}-\xi^{ba\,j}\,\k\iIi aij
-\oh\,\xi^{ac\,i}\,T\Ii b{ac}-\xi^{ba\,i}\,\tau_a\bigr)\,\dx_b\tn\de\yy_i~.$$

\bigbreak
We will be specially involved (\sref{ss:Gauge field theory examples})
in the situation when $\M$ is a pseudo-metric manifold and \hbox{$\xi\equiv{*}\zeta$},
where ${*}$ is the Hodge isomorphism and \hbox{$\zeta:\M\to\OE{r}$}.
Then
$$\xi^{a_1\dots a_r\,i}=
\rdg g^{a_1b_1}\mdots g^{a_rb_r}\,\zeta\iI{b_1\dots b_r}i~.$$

\remark
~The above introduced covariant divergence is somewhat different
from the analogous operation considered in most physics texts,
which is defined as the contraction of the covariant derivation index
with a contravariant index of the tensor acted upon.
For a comparison, consider the interior product
\hbox{$\tilde\xi\equiv(\eta^\#|\xi):\M\to\weu{r}\TO\M\tn\E$},
where $\eta^\#$ is the inverse of a covariantly constant volume form $\eta$\,.
Then we can introduce the covariant divergence $\codiv\tilde\xi$
by interior product in
\hbox{$\nabla\tilde\xi:\M\to\TS\M\tn\weu{r}\TO\M\tn\E$}\,;
a coordinate computation  yields
$$\widetilde{\codiv\xi}=(-1)^{m(r-1)}\,\codiv\tilde\xi~.$$

\subsection{Covariant prolongations in gauge field theory}
\label{ss:Covariant prolongations in gauge field theory}

The geometric setting underlying an essential gauge field theory
is constituted by a vector bundle \hbox{$\E\onto\M$},
possibly with some added fiber structure.
For the moment we are not making any special assumption
about the base manifold $\M$.

The construction of bundles of linear connections of $\E$
can be summarized as follows~\cite{C16c}.
The first jet prolongation \hbox{$\JO\E\onto\E$} is an affine bundle,
while \hbox{$\JO\E\onto\M$} inherits the vector bundle structure.
The affine sub-bundle \hbox{$\C\!\spec{all}\subset\JO\E\tn\E^*$} over $\M$,
constituted by all elements that project onto the identity
\hbox{$\Id{\E}:\M\to\E\tn\E^*$},
can be regarded as the \emph{bundle of all linear connections of $\E$}.
Its `derived' vector bundle is \hbox{$\TS\M\tn\End\E$}.
We also note that \hbox{$\End\E\cong\E\tn\E^*$},
the bundle of all linear fiber endomorphisms of $\E$,
has a Lie-algebra bundle structure determined by the ordinary commutator.

If the fibers of $\E$ are endowed with a more specialized algebraic structure,
then the linear connections preserving it can be regarded as sections
of an affine sub-bundle \hbox{$\C\subset\C\!\spec{all}$}\,,
with derived vector bundle \hbox{$\DO\C=\TS\M\tn\Lie$}
where \hbox{$\Lie\subset\End\E$} is a Lie-subalgebra bundle.
The curvature tensor of a connection can be regarded as a section
\hbox{$\M\to\weu2\TS\M\tn\Lie$}\,.
Linear fiber coordinates $\bigl(\yy^i\bigr)$ on $\E$ also determine
linear fiber coordinates
\hbox{$\bigl(\yy\Ii ij\bigr)\equiv\bigl(\yy^i\tn\yy_j\bigr)$} on $\End\E$.
If we deal with a Lie-subalgebra bundle $\Lie$ then we may choose
`adapted' coordinates
\hbox{$\smash{\bigl(\lfr^\sI\bigr)=\bigl(\lfr^\sI\iI ij\,\yy\Ii ij\bigr)}$},
by which we essentially recover the familiar principal-bundle formalism.
We will use this notation just occasionally, as a shorthand,
but it is easy to realize that all our results
remain valid in a restricted setting anyhow.

\remark
~Any section \hbox{$\xi:\M\to\weu{r}\TS\M\tn\Lie$}
can be also regarded as a basic vertical-valued form
\hbox{$\E\to\weu{r}\TS\M\ten{\E}\VE$}.
Accordingly, its covariant differential $\dO_\k$ with respect to a linear
connection $\k$ could be, in principle, intended in two different ways,
since $\k$ also determines a linear connection of \hbox{$\Lie\onto\M$}.
It is not difficult to check, however, that these two points of view
essentially yield the same $\dO_\k\xi$\,,
with the coordinate expression
$$\dO_\k\xi=\bigl(\de_{a_1}\xi_{a_{2}\dots a_{r+1}}{}^\sI
-[\k_{a_1}\,,\,\xi_{a_2\dots a_{r+1}}]^\sI\bigr)\,
\dx^{a_1}\we\mdots\we\dx^{a_{r+1}}\tn\lfr_\sI~,$$
where the bracket means ordinary commutator.
A similar observation holds about equivalent ways of regarding
a linear connection and its curvature tensor.
\smallbreak

Altogether, the `configuration bundle' of an essential gauge field theory
is \hbox{$\F\equiv\E\cart{\M}\C$},
with sections \hbox{$\phi:\M\to\E$} and sections \hbox{$\k:\M\to\C$}
playing the role of `matter fields' and `gauge fields', respectively.
A couple $(\phi,\k)$ yields the covariant differentials
$$\dO_\k\phi:\M\to\OE{1}~,\qquad \dO_\k\k:\M\to\Omega^2\Lie~,$$
so that we are led to consider the \emph{covariant prolongation bundle}
$$\DF\equiv\F\cart{\M}\OE{1}\cart{\M}\Omega^2\Lie$$
as an analogous of the first jet prolongation $\JO\F$.
We have the morphism
$$\cd:\JO\F\to\DF$$
characterized by
$$\cd\comp(\jO\phi,\jO\k)=(\phi,\k,\dO_\k\phi,\dO_\k\k)~,\qquad
\forall\;(\phi,\k):\M\to\F~,$$
where $\jO\phi$, $\jO\k$ denote the first jet prolongations
of sections.

We denote the induced fiber coordinates on $\C$ by $\bigl(\kk\iIi aij\bigr)$,
and the induced fiber coordinates on \hbox{$\OE{1}\cart{\M}\Omega^2\Lie$} by
\hbox{$\bigl(\xx^a,\yy^i,\zz_a^i\,,\zz\iIi{ab}ij\bigr)$}\,.
We get the coordinate expression
\begin{align*}
&\bigl(\xx^a\,,\,\yy^i\,,\,\kk\iIi aij\,;
\,\zz_a^i\,,\,\zz\iIi{ab}ij\bigr)\comp\cd=
\bigl(\xx^a\,,\,\yy^i\,,\,\kk\iIi aij\,;
\,\yy_a^i\,{-}\,\kk_a\Ii ij\,\yy^j\,,\,
-\kk\iIi ai{j,b}\,{+}\,\kk\iIi bi{j,a}\,{-}\,[\kk_a,\kk_b]\Ii ij\bigr)~,
\\[6pt]
&[\kk_a,\kk_b]\Ii ij\equiv
\kk\iIi aih\,\kk\iIi bhj-\kk\iIi bih\,\kk\iIi ahj~.
\end{align*}

By a straightforward computation one proves:
\begin{proposition}\label{proposition:dHf}
Let \hbox{$f:\DF\to\RR$}\,.
Then there is a unique morphism \hbox{$\dH f:\JO\DF\to\TS\M$} such that
$$\dH(f\comp\cd)=\dH f\comp\JO\cd:\JO_2\F\to\TS\M~,$$
where the left-hand side is the standard horizontal differential
of the function \hbox{$f\comp\cd:\JO\F\to\RR$}
(here $\JO\cd$ is intended as restricted to holonomic jets).
We have the coordinate expression \hbox{$\dH f=\dO_af\,\dx^a$} with
$$\dO_af=\de_af
+(\zz^i_a+\kk\iIi aij\,\yy^j)\,\de_if+\zz^i_{b,a}\,\de^b_if
+\zz\iIi{cb}i{j,a}\,\de\IiI{cb}ij\!f~.$$
\end{proposition}

\section{Covariant-differential Lagrangian field theory}
\label{s:Covariant-differential Lagrangian field theory}
\subsection{Summary of Lagrangian field theory in jet space}
\label{ss:Summary of Lagrangian field theory in jet space}

For a reference,
we sketch the standard geometric formulation of Lagrangian field theory,
which is widely discussed in the literature~%
\cite{Tr67,GS73,Gar74,Kuperschmidt79,HorakKolar83,Kolar84,VinogradovCSS84I_II,
Vitolo91,Costantini94,Krupka02,KrupkaKrupkovaSanders10,Krupka15}.
A \emph{first-order Lagrangian density} on a fibered manifold \hbox{$\E\onto\M$}
is defined to be a totally horizontal $m$-form
$$\Lcal:\JO\E\to\weu{m}\TS\M\subset\weu{m}\TO^*\JO\E~,\qquad m\equiv\dim\M~.$$
We write its coordinate expression as \hbox{$\Lcal=\ell\,\dO^m\xx$}\,,
with $\ell:\JO\E\to\RR$\,.

Consider a morphism
\hbox{$v=v^a\,\de\xx_a+v^i\,\de\yy_i:\JO\E\to\TO\E$} over $\E$\,.
Taking its jet prologation restricted to holonomic jets,
and composing it with the natural morphism \hbox{$\JO\TO\F\to\TO\JO\F$},
we get the natural prolongation~\cite{MaMo83b,C16c}
\begin{align*}
&v_{\sst(1)}=v^a\,\de\xx_a+v^i\,\de\yy_i+v^i_a\,\de\yy^a_i
:\JO_2\E\to\TO\JO\E~,
\\[6pt]
&v^i_a=\dO_av^i-\dO_av^b\,\yy^i_b=
(\de_av^i+\de_jv^i\,\yy^j_a+\de^b_jv^i\,\yy^j_{ab})
-(\de_av^b+\de_jv^b\,\yy^j_a+\de^c_jv^b\,\yy^j_{ac})\,\yy^i_b~.
\end{align*}
Though $v$ is not a vector field on $\JO\E$, we can introduce
a \emph{generalized Lie derivative}
$$\LO_v\Lcal\equiv\dO(v|\Lcal)+v_{\sst(1)}|\dO\Lcal\;:\;\JO_2\E\to\weu{m}\TS\JO\E~.$$
(This kind of order-raising procedure is used in some literature,
in particular by authors working on infinite jets~%
\cite{VinogradovCSS84I_II,Sa89,Vitolo91}.)
We obtain the coordinate expression
$$\LO_v\Lcal=\bigl(\de_a(\ell\,v^a)+\de_i\ell\,v^i
+\de^a_i\ell\,(\dO_av^i-\dO_av^b\,\yy^i_b)\bigr)\,\dO^m\xx
+\ell\,(\de_iv^a\,\dy^i+\de^b_iv^a\,\dy^i_b)\we\dx_a~.$$

We then see that there is a unique morphism
\hbox{$\d_v\Lcal:\JO_2\E\to\weu{m}\TS\M$} over $\M$,
called the \emph{(infinitesimal) variation} of $\Lcal$ determined by $v$\,,
which is characterized by
$$\d_v\Lcal\comp\jO_2\phi=\jO\phi^*\LO_v\Lcal \qquad\forall\phi:\M\to\E~.$$
If $v$ is vertical (\hbox{$v^a=0$}) then we get
\begin{align*}
&\LO_v\Lcal=\d_v\Lcal=(\de_i\ell\,v^i+\de^a_i\ell\,\dO_av^i)\,\dO^m\xx=
\\&\phantom{\LO_v\Lcal=\d_v\Lcal}
=\bigl((\de_i\ell-\dO_a\de^a_i\ell)\,v^i
+\dO_a(v^i\,\de^a_i\ell)\bigr)\,\dO^m\xx\equiv
\\&\phantom{\LO_v\Lcal=\d_v\Lcal}
\equiv\Ecal\pint v+\dH(\DO\Lcal\pint v):\JO_2\E\to\weu{m}\TS\M~,
\end{align*}
where \hbox{$\Ecal:\JO_2\E\to\weu{m}\TS\M\ten{\E}\VO^*\!\E$}
is the Euler-Lagrange operator, and
$$\DO\Lcal=\de^a_i\ell\,\dx_a\tn\dy^i\,:\,
\JO\E\to\weu{m-1}\TS\M\ten{}\VO^*\!\E~,$$
is the fiber derivative of $\Lcal$\,.

We can now express the ``Principle of Least Action'' as the following definition:
a section \hbox{$\phi:\M\to\E$} is called a \emph{critical field}
if for any regular compact subset \hbox{$\K\subset\M$}
and for any vector field \hbox{$v:\E\to\VE$} vanishing on
\hbox{$\cev{\mathsf{p}}(\de\K)\subset\E$} one has
$$\int_\K\d_v\Lcal\comp\jO_2\phi\equiv\int_\K\jO\phi^*\LO_v\Lcal=0~.$$
Critical fields then fulfill the EL field equation
$$\Ecal_i\comp\jO_2\phi\equiv(\de_i\ell-\dO_a\de^a_i\ell)\comp\jO_2\phi=0~.$$

The above introduced fiber derivative $\DO\Lcal$,
via natural operations and identifications,
yields the \emph{momentum} form
$$\Pcal=\Pcal^a_i\,(\dy^i-\yy^i_b\,\dx^b)\we\dx_a\,:\,
\JO\E\to\weu{m}\TO^*\JO\E~,\qquad \Pcal^a_i\equiv\de^a_i\ell~,$$
and the \emph{Poincar\'e-Cartan form}
\hbox{$\Ccal\equiv\Lcal+\Pcal\,:\,\JO\E\to\weu{m}\TO^*\JO\E$}\,.
By straightforward computations one finds
$$\d_v\Pcal=0 \qRq \d_v\Ccal=\d_v\Lcal~.$$

If \hbox{$\phi:\M\to\E$} is any section
then \hbox{$\jO\phi^*(\dy^i-\yy^i_b\,\dx^b)=0$}\,, whence
$$\jO\phi^*(v_{\sst(1)}|\Ccal)=
\bigl[\bigl(\ell\,v^a+
\Pcal^a_i\,(v^i-\yy^i_b\,v^b)\bigr)\comp\jO\phi\bigr]\,\dx_a~.$$
Thus the first-order horizontal form
$$\Jcal\!_v=\bigl(\ell\,v^a+\Pcal^a_i\,(v^i-\yy^i_b\,v^b)\bigr)\,\dx_a:
\JO\E\to\weu{m-1}\TS\M$$
is characterized by
\hbox{$\Jcal\!_v\comp\jO\phi=\jO\phi^*(v_{\sst(1)}|\Ccal)$}
for all sections.

One says that the morphism $v$ is an \emph{on-shell (infinitesimal) symmetry}
of the field theory under consideration if
\hbox{$0=\d_v\Ccal\comp\jO_2\phi=\d_v\Lcal\comp\jO_2\phi$}
for any critical section $\phi$\,.
Moreover, a form \hbox{$\Phi:\JO\E\to\weu{m{-}1}\TO^*\JO\E$}
is called a \emph{conserved current} if
\hbox{$\jO\phi^*\dO\Phi\equiv\dH\Phi\comp\jO_2\phi=0$}
for any critical section $\phi$\,.

By straightforward computations
one finds that the condition that the section \hbox{$\phi:\M\to\E$}
be critical is equivalent to \hbox{$\jO\phi^*(v_{\sst(1)}|\dO\Ccal)=0$}
for any morphism \hbox{$v:\JO\E\to\TO\E$}\,.
The following generalized version of the Noether theorem then holds:
if \hbox{$v:\JO\E\to\TO\E$} is an on-shell symmetry
then \hbox{$v_{\sst(1)}|\Ccal:\JO\E\to\weu{m{-}1}\TO^*\JO\E$}
is a conserved current.
More generally, if there exists an $m\,{-}\,1$-form
\hbox{$\varphi:\JO\E\to\weu{m-1}\TO^*\JO\E$} such that
\hbox{$\jO\phi^*\LO_v\Lcal=\jO\phi^*\dO\varphi$} for all critical sections,
then \hbox{$v_{\sst(1)}|\Ccal\,{-}\,\varphi$} is a conserved current.

\subsection{Covariant Lagrangian density and momentum}
\label{ss:Covariant Lagrangian density and momentum}

According to the \emph{Utiyama principle}%
~\cite{Utiyama56,KolarMichorSlovak93,Janyska07},
in a gauge theory's natural Lagrangian
the matter field's derivatives appear through covariant derivatives,
and a gauge field's derivatives appear through its curvature tensor.
Thus all derivatives appear through the covariant differentials of the fields.

Consider the configuration bundle \hbox{$\F\equiv\E\cart{\M}\C$}
described in~\sref{ss:Covariant prolongations in gauge field theory}.
Then
$$\nabla\!\phi\equiv\dO_\k\phi:\M\to\OE{1}~,\qquad
\rho\equiv-\dO_\k\k:\M\to\Omega^2\Lie~,$$
so that the Lagrangian density can be expressed as a morphism
$$\Lambda\;:\;\DF\equiv\F\cart{\M}\OE{1}\cart{\M}\Omega^2\Lie\to\weu{m}\TS\M~,$$
related to the usual Lagrangian density
\hbox{$\Lcal:\JO\F\equiv\JO\E\cart{\M}\JO\C\to\weu{m}\TS\M$} by
\hbox{$\Lcal=\Lambda\comp\cd$}\,, that is
$$\Lcal\comp(\jO\phi,\jO\k)=\Lambda\comp(\phi,\dO_\k\phi,\dO_\k\k)
\equiv\Lambda\comp\cd\comp(\jO\phi,\jO\k)~.$$
Then we regard $\DF$ as playing a role analogous to that of $\JO\F$
in the standard presentation of Lagrangian field theory,
while \hbox{$(\phi,\k,\dO_\k\phi,\dO_\k\k)$} plays a role analogous
to the first-jet prolongations of sections.
We remark, however, that $\k$ appears in $\Lambda$
only through $\dO_\k\phi$ and $\dO_\k\k$\,.
Thus if we write the local expression \hbox{$\Lambda=\lambda\,\dO^m\xx$}
we have \hbox{$\de\lambda/\de\kk\iIi aij=0$}\,.

We also introduce the ``covariant momentum'' morphism over $\M$
$$\Pi\equiv\bigl(\Pi^{\sst(0)},\Pi^{\sst(1)},\Pi^{\sst(2)}\bigr):
\DF\to\OE{m}^*\cart{\M}\OE{m-1}^*\cart{\M}\Omega^{m-2}\Lie^*$$
obtained by fiber derivative of $\Lambda$ followed by a natural contraction.
Using linear fiber coordinates $\bigl(\yy^i\bigr)$ on $\E$,
and recalling the coordinate notations and conventions introduced
in~\sref{ss:Froelicher-Nijenhuis bracket and covariant differential},
we get
\begin{align*}
&\Pi^{\sst(0)}=\Pi_i\,\dO^m\xx\tn\yy^i~,
&&\Pi^{\sst(1)}=\Pi\Ii ai\,\dx_a\tn\yy^i~,
&&\Pi^{\sst(2)}=\Pi\IiI{ab}ij\,\dx_{ab}\tn\yy\Ii ij~,
\\[6pt]
&\Pi_i\equiv\dde{\lambda}{\yy^i}~,
&&\Pi\Ii ai\equiv\dde{\lambda}{\zz\iI ai}~,
&&\Pi\IiI{ab}ij\equiv\dde{\lambda}{\zz\iIi{ab}ij}~.
\end{align*}
Moreover we have
$$\dO\Lambda=\dO\lambda\we\dO^m\xx=\bigl(\Pi_i\,\dy^i
+\Pi\Ii ai\,\dz\iI ai+\Pi\Ii{ab}\sI\,\dz\iI{ab}\sI\bigr)\we\dO^m\xx~.$$

Setting \hbox{$\ell\equiv\lambda\comp\cd$} we also get
\begin{align*}
&\dde{\ell}{\xx^a}=\dde{\lambda}{\xx^a}\comp\cd~,\qquad
\Pcal_i\equiv\dde{\ell}{\yy^i}=(\Pi_i-\Pi\Ii aj\,\kk\iIi aji)\comp\cd~,\qquad
\Pcal\Ii ai\equiv\dde{\ell}{\yy_a^i}=\Pi\Ii ai\comp\cd~,
\\[6pt]
&\Pcal\IiI aij\equiv\dde{\ell}{\kk\iIi aij}=(-\Pi\Ii ai\,\yy^j
-2\,\Pi\IiI{ab}ih\,\kk\iIi bjh+2\,\Pi\IiI{ab}hj\,\kk\iIi bhi)\comp\cd~,
\\[6pt]
&\Pcal\IiI{b,a}ij\equiv\dde{\ell}{(\kk\iIi aij)_{,b}}=-2\,\Pi\IiI{ab}ij\comp\cd
=2\,\Pi\IiI{ba}ij\comp\cd~.
\end{align*}

\subsection{Variations and field equations}
\label{ss:Variations and field equations}

If \hbox{$v=v^a\,\de\xx_a+v^i\,\de\yy_i+v\iIi aij\,\de\kk\IiI aij\,:
\,\JO\F\to\TO\F$}
then the corresponding infinitesimal variation
of $\Lcal$ is (\sref{ss:Summary of Lagrangian field theory in jet space})
\hbox{$\d_v\Lcal\equiv\d_v\ell\,\dO^m\xx$}\,, with
\begin{align*}
\d_v\ell&=(\Pcal_i-\dO_a\Pcal\Ii ai)\,v^i
+(\Pcal\IiI bij-\dO_a\Pcal\IiI{a,b}ij)\,v\iIi bij
+\dO_a(\Pcal\Ii ai\,v^i+\Pcal\IiI{a,b}ij\,v\iIi bij)+{}
\\&\qquad
+v^a\,\de_a\ell+\ell\,\dO_av^a
-(\yy^i_c\,\Pcal\Ii ai+\kk\iIi bi{j,c}\,\Pcal\IiI {a,b}ij)\,\dO_av^c~.
\end{align*}
Let moreover \hbox{$v=u\comp\cd$} with \hbox{$u:\DF\to\TO\F$}.
Then, recalling proposition~\ref{proposition:dHf}, we obtain
\begin{align*}
\d_v\ell&=\Bigl(
\bigl(\Pi_i-\Pi\Ii aj\,\kk\iIi aji-\dO_a\Pi\Ii ai\bigr)\,u^i
+\bigl(-\Pi\Ii bi\,\yy^j
-2\,\Pi\IiI{bc}ih\,\kk\iIi cjh+2\,\Pi\IiI{bc}hj\,\kk\iIi chi
-2\,\dO_a\Pi\IiI{ab}ij\bigr)\,u\iIi bij+{}
\\
&\qquad
+\dO_a(\Pi\Ii ai\,u^i+2\,\Pi\IiI{ab}ij\,u\iIi bij)
+u^a\,\de_a\lambda+\lambda\,\dO_au^a
-(\yy^i_c\,\Pi\Ii ai+2\,\kk\iIi bi{j,c}\,\Pi\IiI {ab}ij)\,\dO_au^c
\Bigr)\comp\JO\cd
\end{align*}

In the sequel we will denote the covariant prolongation
of \hbox{$(\phi,\k):\M\to\F$} by the shorthand
$$\cd(\phi,\k)\equiv\cd\comp(\jO\phi,\jO\k)\equiv
(\phi,\k,\dO_\k\phi,\dO_\k\k):\M\to\DF~.$$
Evaluating the ``momenta'' \hbox{$\Pi^{\sst(r)}:\DF\to\OE{r}^*$}
through field prolongation we obtain sections
$$\Pi^{\sst(r)}\comp\cd(\phi,\k):\M\to\OE{r}^*~,\qquad r=0,1,2~.$$
We can now apply the results
of~\sref{ss:Froelicher-Nijenhuis bracket and covariant differential}
and of~\sref{ss:Generalized replacement principle} to these objects,
with the only adjustment that, since $\E$ is here replaced by $\E^*$,
their covariant differentials are
$$\dO_\k\bigl(\Pi^{\sst(r)}\comp\cd(\phi,\k)\bigr)=
\fnb{\ost\k\,,\,\Pi^{\sst(r)}\comp\cd(\phi,\k)}
:\M\to\OE{r+1}^*~,$$
where $\ost\k$ is the dual connection of $\k$\,.
We obtain the coordinate expressions
\begin{align*}
\dO_\k(\Pi^{\sst(1)}\comp\cd(\phi,\k))_i&=
(\Pi\Ii aj\,\kk\iIi aji+\dO_a\Pi\Ii ai)\comp\jO\cd(\phi,\k)=
\\&
=(\Pi\Ii aj\comp\cd(\phi,\k))\,\k\iIi aji+\de_a(\Pi\Ii ai\comp\cd(\phi,\k))~,
\\[6pt]
\dO_\k(\Pi^{\sst(2)}\comp\cd(\phi,\k))\IiI{b}ij&=
-2\,\bigl(\dO_a\Pi\IiI{ab}ij
-\kk\iIi ajh\,\Pi\IiI{ab}ih+\kk\iIi ahi\,\Pi\IiI{ab}hj\bigr)\comp\jO\cd(\phi,\k)=
\\&
=-2\,\de_a(\Pi\IiI{ab}ij\comp\cd(\phi,\k))
-2\,\k\iIi ajh\,\Pi\IiI{ab}ih\comp\cd(\phi,\k)
+2\,\k\iIi ahi\,\Pi\IiI{ab}hj\comp\cd(\phi,\k)~,
\end{align*}
which yield a natural interpretation
of the coefficients of $v^i$ and of $v\iIi bij$ in $\d_v\ell$\,.

For the sake of simplicity we will now employ
an abuse of language which is common in physics texts:
when the context is clear we may write $\Pi^{\sst(r)}$
as a shorthand for $\Pi^{\sst(r)}\comp\cd(\phi,\k)$\,.
Similarly we may write
$\Pi\Ii ai$ for $\Pi\Ii ai\comp\cd(\phi,\k)$ and the like.
From the above results we obtain:
\begin{theorem}\label{theorem:fieldequations}
Let \hbox{$(\phi,\k):\M\to\F$}.
The condition that \hbox{$\int_\K\d_v\Lambda\comp\jO\cd(\phi,\k)$}
vanishes for any sufficiently regular compact subset
\hbox{$\K\subset\M$} and for any vertical-valued morphism
\hbox{$v:\JO\F\to\VF$} is equivalent to the validity of the
\emph{field equations}
$$\begin{cases}
\Pi^{\sst(0)}-\dO_\k\Pi^{\sst(1)}=0~,
\\[8pt]
\Pi^{\sst(1)}\tn\phi-\dO_\k\Pi^{\sst(2)}=0~.
\end{cases}$$
\end{theorem}

In simplified coordinate form, the above field equations can be written as
$$\begin{cases}
\Pi_i-\de_a\Pi\Ii ai-\k\iIi aji\,\Pi\Ii aj=0~,
\\[8pt]
\Pi\Ii bi\,\phi^j+2\,(\de_a\Pi\IiI{ab}ij
-\k\iIi ajh\,\Pi\IiI{ab}ih+\k\iIi ahi\,\Pi\IiI{ab}hj)=0~.
\end{cases}$$

\smallbreak\noindent
{\bf Remarks.}\\[2pt]
{\bf a)}~Theorem~\ref{theorem:fieldequations}
may remain valid if its statement is modified by making certain assumptions
about the type of the arbitrary morphism $v$\,,
e.g.\ by restricting it to be a vertical vector field on $\F$
or even a section \hbox{$\M\to\VF$}.
\smallbreak\noindent
{\bf b)}~It is not difficult to check that
the above field equations for the couple $(\phi,\k)$\,,
when written explicitely in coordinates,
do coincide with the Euler-Lagrange field equations
derived from the Lagrangian density $\Lcal$ in the standard theory.
\smallbreak\noindent
{\bf c)}~We can formulate a theory of the field $\phi$ alone,
in which the gauge field $\k$ is a fixed background structure;
then the field equation is just the first of the equations
derived in theorem~\ref{theorem:fieldequations}.

\subsection{Further remarks about variations}
\label{ss:Further remarks about variations}

In the usual formulation of Lagrangian field theory on jet space
(\sref{ss:Summary of Lagrangian field theory in jet space}),
an ``infinitesimal variation'' of the field \hbox{$\phi:\M\to\E$}
can be described as a section \hbox{$\d\phi\equiv v:\M\to\VE$}.
One avails of the natural isomorphism \hbox{$\JO\VE\cong\VO\JO\E$},
which allows the identification of the first jet prolongation
\hbox{$\jO\d\phi:\M\to\JO\VE$} as the variation $\d\jO\phi$\,;
this is needed in the derivation of the usual Euler-Lagrange equations.

Commutation between field variation and field derivation is not valid
when the role of derivation is taken up by the covariant differential,
so we get a slightly more involved situation.
In the context of the previously described essential gauge theory,
an infinitesimal variation of the field $(\phi,\k)$ in the above sense
can be represented as a couple
$$(\d\phi,\d\k):\M\to\E\cart{\M}\Omega^{1}\Lie$$
(since \hbox{$\VE\cong\E\cart{\M}\E$} and \hbox{$\VO\C\cong\C\cart{\M}\Omega^{1}\Lie$}).
It is then natural to set
\begin{align*}
\d\dO_\k\phi&\equiv\d\fnb{\k,\phi}\equiv\fnb{\d\k,\phi}+\fnb{\k,\d\phi}:\M\to\OE{1}~,
\\[6pt]
\d\dO_\k\k&\equiv\d\fnb{\k,\k}\equiv2\,\fnb{\k,\d\k}:\M\to\Omega^{2}\Lie~.
\end{align*}
The reader may wish to compare these expressions to the Lie derivatives
of a linear connection of the tangent bundle of a manifold and its
curvature tensor~\cite{Yano55}.

We can now recover the field equations by a procedure similar to the usual one.
In fact, using again the notational simplification of dropping the explicit
evaluation of the involved objects through the fields,
we get \hbox{$\d\Lambda=\d\lambda\,\dO^m\xx:\M\to\weu{m}\TS\M$} with
\begin{align*}
\d\lambda&=\bigl\langle \dO\lambda\,,\,\d\cd(\phi,\k)\bigr\rangle=
\Pi_i\,\d\phi^i+\Pi\Ii ai\,\fnb{\d\k,\phi}_a^i
+2\,\Pi\IiI{ab}ij\,\fnb{\k,\d\k}\iIi{ab}ij=
\\[6pt]
&=\Pi_i\,\d\phi^i
+\Pi\Ii ai\,(\de_a\d\phi^i-\k\iIi aij\,\d\phi^j-\d\k\iIi aij\,\phi^j)
+2\,\Pi\IiI{ab}ij\,\bigl(\de_a\d\k\iIi{b}ij
+\d\k\iIi{b}ih\,\k\iIi{a}hj-\k\iIi{a}ih\,\d\k\iIi{b}hj\bigr)=
\dbk[6pt]
&=(\Pi_i-\de_a\Pi\Ii ai-\Pi\Ii aj\,\k\iIi aji)\,\d\phi^i
-2\,\bigl(\de_a\Pi\IiI{ab}ij
-\Pi\IiI{ab}ih\,\k\iIi ajh+\Pi\IiI{ab}hj\,\k\iIi ahi
+\oh\,\Pi\Ii bi\,\phi^j\bigr)\,\d\k\iIi{b}ij+{}
\\&\qquad\qquad
+\de_a(\Pi\Ii ai\,\d\phi^i+2\,\Pi\IiI{ab}ij\,\d\k\iIi{b}ij)~.
\end{align*}

\remark
~An \emph{infinitesimal gauge transformation}
can be obtained as a different type of variation,
determined by a section \hbox{$\lfr:\M\to\Lie$}\,.
This can be regarded as a vertical vector field \hbox{$v:\E\to\VE$}
via the rule \hbox{$v(y)\equiv(y,\lfr(y))$}\,, yielding
$$\d\phi\equiv v\comp\phi=\lfr(\phi)~,\qquad
\d\k=\fnb{\k,v}=\fnb{\k,\lfr}~.$$
Clearly this transformation is not suitable for deriving field equations,
as a natural Lagrangian has to be invariant with respect to it.

\subsection{Currents and energy-tensors}
\label{ss:Currents and energy-tensors}

The Poincar\'e-Cartan form for an essential gauge field theory
of matter and gauge fields is
\hbox{$\Ccal\equiv\Lcal+\Pcal:\JO\F\to\weu{m}\TO^*\JO\F$} where
$$\Pcal=\bigl(\Pcal^a_i\,(\dy^i-\yy^i_b\,\dx^b)
+\Pcal\IiI{a,b}ij\,(\dO\kk\iIi bij-\kk\iIi bi{j,c}\,\dx^c)
\bigr)\we\dx_a~.$$
It is apparent that, in general, $\Ccal$ cannot be completely recovered
only in terms of covariant prolongations.
Nevertheless, the current associated with a morphism
\hbox{$v:\JO\F\to\TO\F$}
can be actually seen as an object ``living'' on $\DF$,
provided that $v$ is of a suitably restricted type.

Let \hbox{$v=u\comp\cd:\JO\F\to\TO\F$} with \hbox{$u:\DF\to\TO\F$}.
We have the current associated with $v$\,, that is the horizontal form
(\sref{ss:Summary of Lagrangian field theory in jet space})
\begin{align*}
&\Jcal\!_v=\Jcal^a\,\dx_a:\JO\F\to\weu{m-1}\TS\M~,
\\[6pt]
&\Jcal^a=\ell\,v^a+\Pcal^a_i\,(v^i-\yy^i_b\,v^b)
+\Pcal\IiI{a,b}ij\,(v\iIi bij-\kk\iIi bi{j,c}\,v^c)~.
\end{align*}

We would like to write \hbox{$\Jcal\!_v=\Ical\!_u\comp\cd$}
with \hbox{$\Ical\!_u:\DF\to\weu{m-1}\TS\M$}.
Since
$$\Jcal^a=\bigl(\lambda\,u^a
+\Pi^a_i\,(u^i-(\zz^i_b+\kk\iIi bij\,\yy^j)\,u^b)
+2\,\Pi\IiI{ab}ij\,(u\iIi bij-\kk\iIi bi{j,c}\,u^c)\bigr)\comp\cd~,$$
we see that the obstruction to doing so
lies in the need for expressing $\kk\iIi bi{j,c}$ via a function on $\DF$;
the obstruction disappears, in particular, when $u$ is vertical-valued.

Another special case is that of the ``horizontal lift''
of a vector field \hbox{$\ul u=u^a\,\de\xx_a$} on $\M$.
Actually a natural generalization of the usual notion of horizontal lift
of a basic vector field via a connection,
exploiting the notion of ``overconnection''~\cite{C16c}, yields the morphism
$$\ul u^\up:\E\cart{\M}\JO\C\to\TO\F$$
with the coordinate expression
$$\ul u^\up=u^a\,\bigl(\de\xx_a+\kk\iIi aij\,\yy^j\,\de\yy_i
+(\kk\iIi ai{j,b}-\kk\iIi ahj\,\kk^i_{bh}+\kk^h_{bj}\,\kk\iIi aih)\,
\de\kk\iIi bij \bigr)~.$$
Inserting $\ul u^\up$ into $\Jcal^a$ in the place of $v$,
by a short computation we obtain
$$\Jcal^a=u^b\,\bigl(\lambda\,\d\Ii ab-\Pi^a_i\,\zz_b^i
-2\,\Pi\IiI{ac}ij\,\zz\iIi{bc}ij\bigr)\comp\cd~.$$

Summarizing, we can express the above discussion as follows.
\begin{proposition}
Let \hbox{$\ul u:\M\to\TO\M$},
\hbox{$w=w^i\,\de\yy_i+w\iIi aij\,\de\kk\IiI aij:\DF\to\VO\F$}, and set
$$v\equiv \ul u^\up+w\comp\cd:\JO\F\to\TO\F~.$$
Then there exists \hbox{$\Ical_{\ul u,w}:\DF\to\weu{m-1}\TS\M$}
such that \hbox{$\Jcal\!_v=\Ical_{\ul u,w}\comp\cd$}\,, with the components
$$\Ical^a=u^b\,\bigl(\lambda\,\d\Ii ab-\Pi^a_i\,\zz_b^i
-2\,\Pi\IiI{ac}ij\,\zz\iIi{bc}ij\bigr)
+\Pi^a_i\,w^i+2\,\Pi\IiI{ab}ij\,w\iIi bij~.$$
\end{proposition}

The notion of ``canonical energy-tensor''%
~\cite{FernandezGarciaRodrigo00,FerrarisFrancaviglia92,
FatibeneFerrarisFrancaviglia94,ForgerRoemer04,Leclerc06,ObukhovPuetzfeld14,Pons09}
is essentially about relating conserved currents
with vector fields on the base manifold.
Though the expression \hbox{$\ell\,\d^a_b-\phi^i_{,a}\,\de^a_i\ell$}
found in the literature is usually recognized to be devoid of
geometric meaning in general,
its truly covariant modification introduced by Hermann~\cite{Hermann75}
is still not very well-known.
The construction requires a connection of the theory's configuration bundle,
as this is the most natural way to lift basic vector fields;
in terms of the tensor's coordinate expression, it amounts to replacing
the field's ordinary derivative $\phi^i_{,a}$ with its covariant derivative
with respect to the assumed connection~\cite{CM85,C16c}.

The Hermann construction is nicely suited to be extended
to a theory of coupled matter and gauge fields $(\phi,\k)$.
In this case no extra structure is needed,
as one avails of the covariant differentials $(\dO_\k\phi,\dO_\k\k)$\,;
one obtains the joint canonical energy-tensor
$$\Ucal:\JO\F\to\weu{m-1}\TS\M\tn\TS\M~,$$
which has the coordinate expression
\begin{align*}
\Ucal\Ii ab&=\ell\,\d\Ii ab-\Pcal\Ii ai\,(\yy^i_b-\kk\iIi bij\,\yy^j)
-\Pcal\IiI{a,c}ij\,(
-\kk\iIi ai{j,b}\,{+}\,\kk\iIi bi{j,a}\,{-}\,[\kk_a,\kk_b]\Ii ij)=
\\[6pt]
&=\bigl(\lambda\,\d\Ii ab
-\Pi\Ii ai\,\zz^i_b-2\,\Pi\IiI{ac}ij\,\zz\iIi{ab}ij\bigr)\comp\cd\equiv
\Upsilon\Ii ab\comp\cd~.
\end{align*}

We then see that there exists a unique ``covariant canonical energy-tensor'' 
$$\Upsilon:\DF\to\weu{m-1}\TS\M\tn\TS\M$$
such that \hbox{$\Ucal=\Upsilon\comp\cd$}\,.
Moreover for any basic vector field $\ul u$ we have
\hbox{$\Ical_{\ul u}=\Upsilon\pint\ul u$}\,.

\section{Field theory in spacetime}
\label{s:Field theory in spacetime}
\subsection{Gauge field theory on a General Relativistic background}
\label{ss:Gauge field theory on a General Relativistic background}

We consider an essential gauge field theory setting as
in~\sref{ss:Covariant Lagrangian density and momentum},
but now the base manifold $\M$ is assumed to be an oriented Lorentz spacetime
(\hbox{$m=4$}).
We will denote the metric, the spacetime connection and the unit volume form
as $g$\,, $\sGa$ and $\eta$\,, respectively.

We allow the matter field $\phi$ to have spacetime indices,
namely the interacting matter and gauge fields constitute a section
$$(\phi,\k):\M\to\F\equiv(\E\tn\Y)\cart{\M}\C$$
where \hbox{$\Y\subset
(\mathop{\textstyle\otimes}\!\TO\M)\tn(\mathop{\textstyle\otimes}\!\TS\M)$}
is a vector sub-bundle of the tensor algebra of $\TO\M$.
Thus $\phi$ can be regarded as a charged bosonic field
of possibly non-zero spin
(in~\sref{ss:Gauge field theory examples} we will also consider fermionic fields).

We indicate fiber coordinates of $\E\tn\Y$ as
$\yy^{i\sA}$, where ${\scriptstyle A}$
represents the appropriate set of spacetime indices.
Accordingly, the coefficients of the connection of $\Y$
determined by the spacetime connection will be denoted as $\G\!\iIi a\sA\sB$\,.
We have the covariant differential
$$\nabla\!\phi\equiv\dO_{\k\sst{\otimes}\G}\phi\equiv\fnb{\k\tn\G\,,\,\phi}$$
where the ``tensor product connection'' $\k\tn\G$
is the induced connection of $\E\tn\Y$, namely
$$\nabla\!_a\phi^{i\sA}=\de_a\phi^{i\sA}-\k\iIi aij\,\phi^{j\sA}
-\G\!\iIi a\sA\sB\,\phi^{i\sB}~,$$
and the field \hbox{$(\phi,\k):\M\to\F$} has the covariant prolongation
$$\cd(\phi,\k)\equiv\cd\comp(\jO\phi,\jO\k)\equiv
\bigl(\dO_{\k\sst{\otimes}\G}\phi\,,\,\dO_\k\k\bigr)~.$$

The basic setting laid out in~\sref{ss:Covariant Lagrangian density and momentum}
must be now adjusted.
On \hbox{$\DF\onto\M$} we have fiber coordinates
$\bigl(\yy^{i\sA},\kk\iIi aij\,,\zz\iI a{i\sA},\zz\iIi{ab}ij\bigr)$.
The ``covariant'' momentum morphism
$$\Pi\equiv\bigl(\Pi^{\sst(0)},\Pi^{\sst(1)},\Pi^{\sst(2)}\bigr):
\DF\to(\OE{m}^*{\otimes}\Y^*)\cart{\M}(\OE{m-1}^*{\otimes}\Y^*)
\cart{\M}\Omega^{m-2}\Lie^*$$
has components
$\bigl(\Pi_{i\sA}\,,\,\Pi\Ii a{i\sA}\,,\,\Pi\IiI{ab}ij\,\bigr)$,
and the components of the standard momentum $\Pcal$ are related to these by
$$\Pcal_{i\sA}=\bigl(\Pi_{i\sA}-\Pi\Ii a{j\sB}\,
(\kk\iIi aji\,\d\Ii\sB\sA+\d\Ii ji\,\G\!\iIi a\sB\sA)\,\bigr)\comp\cd~,\qquad
\Pcal\Ii a{i\sA}=\Pi\Ii a{i\sA}\comp\cd~,\qquad
\Pcal\IiI{b,a}ij=2\,\Pi\IiI{ba}ij\comp\cd~.$$

The procedure which led to theorem~\ref{theorem:fieldequations}
can be adapted to this situation without difficulty,
and employing again the notational simplification used there|%
$\Pi^{\sst(r)}$ for $\Pi^{\sst(r)}\comp\cd(\phi,\k)$
and the like for the momentum components|we find the field equations
$$\begin{cases}
\Pi^{\sst(0)}-\dO_{\k\sst{\otimes}\G}\Pi^{\sst(1)}=0~,
\\[8pt]
\Pi^{\sst(1)}\tn\phi-\dO_\k\Pi^{\sst(2)}=0~,
\end{cases}$$
that is, in coordinate form,
$$\begin{cases}
\Pi_{i\sA}-\Pi\Ii a{j\sB}\,
(\k\iIi aji\,\d\Ii\sB\sA+\d\Ii ji\,\G\!\iIi a\sB\sA)
-\de_a\Pi\Ii a{i\sA}=0~,\\[8pt]
\Pi\Ii b{i\sA}\,\phi^{j\sA}+2\,(\de_a\Pi\IiI{ab}ij
+\Pi\IiI{ab}hj\,\k\iIi ahi-\Pi\IiI{ab}ih\,\k\iIi ajh)=0~.
\end{cases}$$

\remark
~By contraction with the inverse $\eta^\#$ of $\eta$ we get
(\sref{ss:Generalized replacement principle}) the ``contravariant momentum''
$$\tilde\Pi\equiv
\bigl(\tilde\Pi^{\sst(0)},\tilde\Pi^{\sst(1)},\tilde\Pi^{\sst(2)}\bigr):
\DF\to(\E^*\tn\Y^*)\cart{\M}(\TO\M\tn\E^*\tn\Y^*)\cart{\M}(\weu2\TO\M\tn\Lie^*)~,$$
with the components
$$\tilde\Pi_{i\sA}=\Pi_{i\sA}/\rdg~,\qquad
\tilde\Pi\Ii a{i\sA}=\Pi\Ii a{i\sA}/\rdg~,\qquad
\tilde\Pi\IiI{ab}ij=\Pi\IiI{ab}ij/\rdg~.$$
The replacement principle for $\Pi^{\sst(1)}$ holds, in the present case,
in the modified form
\begin{align*}
(\codiv\tilde\Pi^{\sst(1)})_{i\sA}&=
\tfrac1\rrdg\,(\codiv\Pi^{\sst(1)})_{i\sA}=
\tfrac1\rrdg\,(\dO_{\k\sst{\otimes}\G}\Pi^{\sst(1)}-\tau\we\Pi^{\sst(1)})_{i\sA}=
\\&
=\tfrac1\rrdg\,(\de_a\Pi\Ii a{i\sA}
+\k\iIi aji\,\Pi\Ii a{j\sA}+\G\!\iIi a\sB\sA\,\Pi\Ii a{i\sB}-\tau_a\,\Pi\Ii ai)~,
\end{align*}
while it is unchanged for $\Pi^{\sst(2)}$\,.
Hence the field equations can be recast
in terms of covariant divergences in the form
$$\begin{cases}
\tilde\Pi^{\sst(0)}-\codiv\tilde\Pi^{\sst(1)}=\text{torsion terms}~,
\\[8pt]
\tilde\Pi^{\sst(1)}\tn\phi-\codiv\tilde\Pi^{\sst(2)}=\text{torsion terms}~.
\end{cases}$$

\subsection{Stress-energy tensor}
\label{ss:Stress-energy tensor}

We now revisit, in the context introduced
in~\sref{ss:Gauge field theory on a General Relativistic background},
the usual approach to stress-energy tensors in Einstein spacetime.
This notion develops from considering infinitesimal deformations
determined by a  ``basic'' vector field \hbox{$\ul u:\M\to\TO\M$}\,,
so it should be related to the procedure yielding
the ``canonical'' energy-tensor (\sref{ss:Currents and energy-tensors});
it is actually well-known~\cite{GotayMarsden92}
that in standard theories the two objects are nearly the same.

The argument under consideration is based
on the notion of Lie derivative of the fields with respect to $\ul u$\,;
the usual formulation~\cite{LandauLifchitz68,HE73}
explicitely considers fields with spacetime indices only.
If the fields also have indices of other kinds then their Lie derivatives
are not well-defined in general,
but the argument can be successfully carried on by employing
a suitable extension.

One such extension can be introduced in terms
of a possibly local, linear ``reference connection'' $\varkappa$ of $\E$.
For any \hbox{$\xi:\E\to\weu{r}\TS\M\ten{\E}\TO\E$} we consider
the ``covariant Lie derivative''
$$\d\xi\equiv\LO_{\varkappa,\ul u}\xi\equiv
i_{\ul u}\dO_{\varkappa}\xi+\dO_{\varkappa}(i_{\ul u}\xi):
\E\to\weu{r+1}\TS\M\ten{\E}\VE~.$$
Then by comparing coordinate expressions it is not difficult to show that
$$\LO_{\varkappa,\ul u}\xi=
(\LO_{\ul u\pint\varkappa}\xi)\pint\omega_\varkappa~,$$
where $\LO_{\ul u\pint\varkappa}\xi$ is the ordinary Lie derivative
of $\xi$\,, seen as a tensor field on $\E$, with respect to the
horizontal lift \hbox{$\ul u\pint\varkappa:\E\to\TO\E$},
and \hbox{$\omega_\varkappa:\TO\E\to\VE$} is the vertical projection form
associated with $\varkappa$\,.
(See Kola\v{r}-Michor-Slovak~\cite{KolarMichorSlovak93} for a discussion
of Lie derivatives from a general point of view.)

For \hbox{$\xi:\M\to\OE{r}$} we then get
\hbox{$\d\xi:\M\to\OE{r}$}.
If \hbox{$\sigma:\M\to\E$} is a section and $\k$ is a linear connection of $\E$
then we easily find
$$\d\dO_\k\sigma=\dO_\k\d\sigma-(\d\k)\pint\sigma~.$$
For \hbox{$t:\M\to\Y$} let us now define \hbox{$\d t\equiv\LO_{\ul u}t$}
to be the standard Lie derivative.
Then it is not difficult to see that the operator $\d$
can be naturally extended, via linearity and the Leibnitz rule,
to act on sections \hbox{$\phi:\M\to\E\tn\Y$}.
Similarly we set
$$\d(\k\tn\G)\equiv\d\k\tn\G+\k\tn\d\G~,$$
where \hbox{$\d\G\equiv\LO_{\ul u}\G$} is the standard Lie derivative
of the (tensor extension of the) spacetime connection~\cite{Yano55}.
We obtain
\begin{align*}
&\d\nabla\!\phi=\nabla\d\phi-\d(\k\tn\G)\pint\phi\quad\text{i.e.}\quad
(\d\nabla\!\phi)\iI a{i\sA}=
\nabla\!_a\d\phi^{i\sA}-\d\k\iIi aij\,\phi^{j\sA}-\d\G\!\iIi a\sA\sB\,\phi^{i\sB}~,
\\[6pt]
&\d\fnb{\k,\k}=2\,\fnb{\k,\d\k}~,\qquad \d\fnb{\G,\G}=2\,\fnb{\G,\d\G}~,
\\[6pt]
&\d\fnb{\k\tn\G,\k\tn\G}\iIi{ab}{i\sA}{j\sB}=
2\,\fnb{\k,\d\k}\iIi{ab}ij\,\d\Ii\sA\sB
+2\,\d\Ii ij\,\fnb{\G,\d\G}\iIi{ab}\sA\sB~,
\\[6pt]
&\fnb{\k,\d\k}\iI{ab}i=
\de_a\d\k\iIi bij-\k\iIi aih\,\d\k\iIi bhj+\d\k\iIi bih\,\k\iIi ahj~.
\end{align*}

\remark
~We may choose $\varkappa$ to be curvature-free (a local gauge);
then we may work in local charts such that the coefficients of $\varkappa$ vanish,
getting
$$\d s^i=u^a\,\de_as^i~,\qquad
\d\k\iIi bij=\de_bu^a\,\k\iIi aij+u^a\,\de_a\k\iIi bi{j}~.$$
Such choice is then equivalent to using standard expressions for Lie derivatives
by ignoring the fiber indices of $\E$\,;
namely, these indices are now seen as mere labels,
so that e.g.\ $\bigl(s^i\bigr)$ is treated as a collection of scalar functions
and $\bigl(\k\iIi bij\bigr)$ is treated as a collection of 1-forms on $\M$.
Or, we could use $\k$ itself as the reference connection, obtaining in particular
\hbox{$\d\k=i_{\ul u}\dO_\k\k$}\,.
\smallbreak

Expressing the Lagrangian density as \hbox{$\Lambda=\tilde\lambda\,\eta$}
we get
\begin{align*}
\d\Lambda&=(\d\tilde\lambda)\,\eta+\tilde\lambda\,\d\eta~,\qquad
\d\eta=\oh\,g^{ab}\,\d g_{ab}\,\eta
\quad(\text{with}~\d g_{ab}\equiv\LO_{\ul u}g_{ab})~,
\\[8pt]
\d\tilde\lambda&=\tfrac{\de\tilde\lambda}{\de g_{ab}}\,\d g_{ab}
+\tilde\Pi_{i\sA}\,\d\phi^{i\sA}
+\tilde\Pi\Ii a{i\sA}\,(\d\nabla\!\phi)\iI a{i\sA}
+2\,\tilde\Pi\IiI{ab}ij\,\fnb{\k,\d\k}\,\iIi{ab}ij=
\\[6pt]
&=\tfrac{\de\tilde\lambda}{\de g_{ab}}\,\d g_{ab}
-\tilde\Pi\Ii a{i\sA}\,\d\G\!\iIi a\sA\sB\,\phi^{i\sB}+{}
\\&\qquad
+\bigl(\tilde\Pi_{i\sA}-\nabla\!_a\tilde\Pi\Ii a{i\sA}\bigr)\,\d\phi^{i\sA}
+\nabla\!_a(\tilde\Pi\Ii a{i\sA}\,\d\phi^{i\sA})
+2\,\nabla\!_a(\tilde\Pi\IiI{ab}ij\,\d\k\iIi bij)+{}
\\&\qquad
+\bigl(
-\tilde\Pi\Ii b{i\sA}\,\phi^{j\sA}
-2\,\de_a\tilde\Pi\IiI{ab}ij+\G\!\iIi acc\,\tilde\Pi\IiI{ab}ij
-2\,\k\iIi ahi\,\tilde\Pi\IiI{ab}hj
+2\,\tilde\Pi\IiI{ab}ih\,\k\iIi ajh
\bigr)\,\d\k\iIi bij~,
\end{align*}
where we assumed that $\tilde\lambda$ depends on the base coordinates
only through the metric and its derivatives.

We now make the further assumption that $\G$ is the the Levi-Civita connection,
namely it is torsionless in addition to being metric.
Then by standard computations one shows that
the second term in the above expression of $\d\tilde\lambda$
can be expressed in the form
$$-\tilde\Pi\Ii a{i\sA}\,\d\G\!\iIi a\sA\sB\,\phi^{i\sB}=
\Scal^{ab}\,\d g_{ab}+\text{a divergence}~,$$
where $\Scal$ is a symmetric tensor field.
Moreover we observe that the vanishing of the torsion,
taking the replacement principle (\sref{ss:Generalized replacement principle})
into account,
implies that the coefficients of $\d\phi^{i\sA}$ and $\d\k\iIi bij$
in $\d\tilde\lambda$ vanish when $(\phi,\k)$ obeys the field equations.
Hence, picking out those terms in $\d\Lambda$ that are not divergences
and do not vanish on-shell, we prove:
\begin{theorem}\label{theorem:stress-energy_tensor}
Let $(\M,g)$ be a Lorentz spacetime and $\G$ the related Levi-Civita connection.
Let \hbox{$\Lambda=\tilde\lambda\,\eta:\DF\to\weu4\TS\M$} be such that
$\tilde\lambda$ explicitely depends on spacetime coordinates
only through $g$ and its derivatives.
There there exists a unique morphism
$$\Tcal:\JO\DF\to\TS\M\tn\TS\M~,$$
called the \emph{stress-energy tensor},
with the following property:
if the vector field \hbox{$\ul u:\M\to\TO\M$} vanishes on the boundary
of the compact subset \hbox{$\D\subset\M$}
and $(\phi,\k)$ obeys the field equations, then
$$\int_\D\d\Lambda\comp\cd(\phi,\k)=
\int_\D (\Tcal^{ab}\comp\jO\cd(\phi,\k))\,\d g_{ab}\,\eta~.$$
Moreover $\Tcal$ turns out to be symmetric,
as we obtain the coordinate expression
$$\Tcal^{ab}=
\tfrac{\de\tilde\lambda}{\de g_{ab}}+\oh\,\tilde\lambda\,g^{ab}+\Scal^{ab}~,$$
where $\Scal$ is the symmetric tensor arising from the
dependence of $\Lambda$ from the derivatives of $g$
through the spacetime connection.
\end{theorem}

By a standard argument then one also proves:
\begin{theorem}\label{theorem:divergence-free_stress-energy_tensor}
The stress-energy tensor is ``on-shell'' divergence-free,
namely for any critical field \hbox{$(\phi,\k):\M\to\F$} we have
$$\codiv(\Tcal\comp\jO\cd(\phi,\k))=0~.$$
\end{theorem}

\subsection{Gauge field theory examples}
\label{ss:Gauge field theory examples}

We work out the basic examples, in the covariant differential setting,
of a gauge field interacting with either a boson field or a Dirac field.
The gauge Lagrangian is defined to be
\hbox{$\Lambda\spec{gauge}=\lambda\spec{gauge}\,\dO^4\xx$} with
$$\lambda\spec{gauge}=
\oq\,g^{ac}\,g^{bd}\,\zz\iIi{ab}ij\,\zz\iIi{cd}ji\,\rdg~.$$

\subsubsection*{Bosonic field}
As for the matter field we consider a complication
with respect to the general scheme,
namely it now consists of a couple
$$(\phi,\bar\phi):\M\to(\E\tn\Y)\cart{\M}(\E^*\tn\Y^*)$$
of \emph{mutually independent} fields valued into mutually dual bundles.
In the usual formulations $\phi$ and $\bar\phi$ are often regarded
as mutually adjoint fields through some Hermitan fiber structure,
but that particularization is not needed here.

It is not difficult to see
that the field equations (theorem~\ref{theorem:fieldequations})
must now be rewritten in the adapted form
$$\begin{cases}
\Pi^{\sst(0)}-\dO_{\k\sst{\otimes}\G}\Pi^{\sst(1)}=0~,
\\[6pt]
\bar\Pi^{\sst(0)}-\dO_{\k\sst{\otimes}\G}\bar\Pi^{\sst(1)}=0~,
\\[6pt]
\Pi^{\sst(1)}\tn\phi-\bar\phi\tn\bar\Pi^{\sst(1)}-\dO_\k\Pi^{\sst(2)}=0~,
\end{cases}$$
that is
$$\begin{cases}
\Pi_{i\sA}-\Pi\Ii a{j\sB}\,
(\k\iIi aji\,\d\Ii\sB\sA+\d\Ii ji\,\G\!\iIi a\sB\sA)
-\dO_a\Pi\Ii a{i\sA}=0~,\\[8pt]
\Pi^{i\sA}+\Pi^{aj\sB}\,
(\k\iIi aij\,\d\Ii\sA\sB+\d\Ii ij\,\G\!\iIi a\sA\sB)
-\dO_a\Pi^{ai\sA}=0~,\\[8pt]
\Pi\Ii b{j\sA}\,\phi^{i\sA}-\Pi^{bi\sA}\,\bar\phi_{j\sA}
+2\,(\de_a\Pi\IiI{ab}ij
+\Pi\IiI{ab}hi\,\k\iIi ahj-\Pi\IiI{ab}jh\,\kk\iIi aih)=0~,
\end{cases}$$
where $\bar\Pi^{\sst(0)}$ and $\bar\Pi^{\sst(1)}$,
with components $\Pi^{i\sA}$ and $\Pi^{ai\sA}$,
denote the momenta related to the dual sector,
and compositions of the momenta by $\cd(\phi,\bar\phi,\k)$ are intended.
Note that the same gauge field interacts with $\phi$ and $\bar\phi$\,.
On the other hand we could consider independent gauge fields
$\k$ and $\bar\k$\,, getting one more field equation;
then by identifying $\bar\k$ as the dual of $\k$
(\hbox{$\bar\k\iI{ai}j=-\k\iIi aji$})
we obtain the above field equation for $\k$\,.

We set \hbox{$\Lambda\equiv\Lambda\spec{matter}+\Lambda\spec{gauge}$}
with \hbox{$\Lambda\spec{matter}=\lambda\spec{boson}\,\dO^4\xx$} and
$$\lambda\spec{boson}=
\oh\,(g^{ab}\,\zz_{ai\sA}\,\zz_b^{i\sA}-m^2\,\yy_{i\sA}\,\yy^{i\sA})~.$$

The explicit derivation of the field equations is now a straightforward task
(maybe somewhat simpler than their usual derivation
as the Euler-Lagrange equations).
The momenta can be immediately expressed in coordinate-free form as
\begin{align*}
&\Pi^{\sst(0)}=-\oh\,m^2\,\eta\tn\bar\phi~,
&&\bar\Pi^{\sst(0)}=-\oh\,m^2\,\eta\tn\phi~,
\\[6pt]
&\Pi^{\sst(1)}=\oh\,{*}\dO_{\k\sst{\otimes}\G}\bar\phi\equiv
\oh\,{*}\nabla\!\bar\phi~,
&&\bar\Pi^{\sst(1)}=\oh\,{*}\dO_{\k\sst{\otimes}\G}\phi\equiv
\oh\,{*}\nabla\!\phi~,
\\[6pt]
&\Pi^{\sst(2)}=\oh\,{*}\dO_\k\k~,
\end{align*}
where ${*}$ denotes the Hodge isomorphism.
Hence the field equations can be cast (up to obvious transpositions)
in the coordinate-free form
$$\begin{cases}
\nabla({*}\nabla\!\bar\phi)+m^2\,\eta\tn\bar\phi=0~,
\\[6pt]
\nabla({*}\nabla\!\phi)+m^2\,\eta\tn\phi=0~,
\\[6pt]
({*}\nabla\!\bar\phi)\tn\phi-\bar\phi\tn({*}\nabla\!\phi)
-\dO_\k{*}\dO_\k\k=0~.
\end{cases}$$

Using shorthands \hbox{$\rho\equiv-\dO_\k\k$}\,,
\hbox{$\rho\Ii{abi}j\equiv g^{ac}\,g^{bd}\,\rho\iIi{cd}ij$}\,,
we find the coordinate expressions
$$\begin{cases}
\tfrac1\rrdg\,\de_a(g^{ab}\,\nabla\!_b\bar\phi_{i\sA}\,\rdg)+m^2\,\bar\phi_{i\sA}
+g^{ab}\,\nabla\!_b\bar\phi_{j\sB}\,
(\k\iIi aji\,\d\Ii\sB\sA+\d\Ii ji\,\G\!\iIi a\sB\sA)=0~,
\\[8pt]
\tfrac1\rrdg\,\de_a(g^{ab}\,\nabla\!_b\phi^{i\sA}\,\rdg)+m^2\,\phi^{i\sA}
-g^{ab}\,\nabla\!_b\phi^{jsB}\,(\k\iIi aij\,\d\Ii\sA\sB+\d\Ii ij\,\G\!\iIi a\sA\sB)=0~,
\\[8pt]
\tfrac1\rrdg\,\de_a\bigl(\rho\Ii{abi}j\,\rdg\bigr)
+\rho\Ii{abi}h\,\k\iIi ahj-\k\iIi aih\,\rho\Ii{abh}j
+\oh\,g^{ab}\,(\bar\phi_j\,\nabla\!_a\phi^i-\nabla\!_a\bar\phi_j\,\phi^i)=0~.
\end{cases}$$

By virtue of the replacement principle (\sref{ss:Generalized replacement principle}),
the field equations can also be written in terms of covariant divergences.

\subsubsection*{Spin-$\oh$ field}
The geometric setting for Dirac spinors in curved spacetime
has finer points, not examined here, that are widely discussed in the literature.
My own view about this subject has been expressed
in previous papers~\cite{C07,C16c}.

The \emph{4-spinor} bundle \hbox{$\W\onto\M$} is endowed with
a linear morphism \hbox{$\g:\TO\M\to\End\W$} whose components
(the ``gamma matrices'') are constant in a suitable frame.
Allowing further internal degrees of freedom besides spin,
the matter field can be described as a section
$$(\psi,\bar\psi):\M\to(\W\tn\E)\cart{\M}(\W^*\tn\E^*)~.$$
Besides the gauge field we have to deal with a \emph{spinor connection} $\Cs$,
that is a linear connection of $\W$;
in the present context it is considered as a fixed structure,
related to the gravitational background.
The tensor product connection $\Cs\tn\k$ of $\W\tn\E$ has then the components
$$(\Cs\tn\k)\iIi a{\a i}{\b j}=
\Cs\!\iIi a\a\b\,\d\Ii ij+\d\Ii\a\b\,\kk\iIi aij~.$$

The field equations are now
$$\begin{cases}
\Pi^{\sst(0)}-\fnb{\Cs\tn\k\,,\,\Pi^{\sst(1)}}=0~,
\\[6pt]
\bar\Pi^{\sst(0)}-\fnb{\Cs\tn\k\,,\,\bar\Pi^{\sst(1)}}=0~,
\\[6pt]
\Pi^{\sst(1)}\tn\psi-\bar\psi\tn\bar\Pi^{\sst(1)}-\fnb{\k\,,\,\Pi^{\sst(2)}}=0~,
\end{cases}$$
that is
$$\begin{cases}
\Pi_{\a i}-\Pi\Ii a{\b j}\,(\Cs\tn\k)\iIi a{\b j}{\a i}
-\de_a\Pi\Ii a{\a i}=0~,\\[8pt]
\Pi^{\a i}+\Pi^{a\b j}\,(\Cs\tn\k)\iIi a{\a i}{\b j}
-\de_a\Pi^{a\a i}=0~,\\[8pt]
\Pi\Ii b{\a j}\,\psi^{\a i}-\bar\psi_{\a j}\,\Pi^{b\a i}
+2\,(\de_a\Pi\IiI{ab}ji
+\Pi\IiI{ab}hi\,\k\iIi ahj-\Pi\IiI{ab}jh\,\k\iIi aih)=0~.
\end{cases}$$
Comparing these
to the generic field equations in a gravitational background
(\sref{ss:Gauge field theory on a General Relativistic background})
one notes that here the spinor indices take up the role
of the spacetime indices there.

The gauge sector Lagrangian is the same as in the boson case.
The matter Lagrangian
is the Dirac Lagrangian \hbox{$\lambda\spec{Dirac}\,\dO^4\xx$}\,,
with
$$\lambda\spec{Dirac}=\bigl(\ih\,g^{ab}\,(\yy_{\a i}\,\g\iIi a\a\b\,\zz\iI b{\b i}
-\zz_{a\a i}\,\g\iIi b\a\b\,\yy^{\b i})
-m\,\yy_{\a i}\,\yy^{\a i}\bigr)\,\rdg~,$$
whence we get
\begin{align*}
&\Pi^{\sst(0)}=
(-\ih\,\nasl\bar\psi-m\bar\psi)\tn\eta\equiv
{*}(-\ih\,\nasl\bar\psi-m\bar\psi)~,
\\[6pt]
&\bar\Pi^{\sst(0)}=(\ih\,\nasl\psi-m\psi)\tn\eta\equiv
{*}(\ih\,\nasl\psi-m\psi)~,
\\[6pt]
&\Pi^{\sst(1)}=\ih\,{*}(\bar\psi\,\g)~,\qquad
\bar\Pi^{\sst(1)}=-\ih\,{*}(\g\,\psi)~.
\end{align*}

The field equations for the matter field then become
$$\begin{cases}
\ih\,\fnb{\Cs\tn\k\,,\,{*}(\bar\psi\,\g)}+{*}(\ih\,\nasl\bar\psi+m\bar\psi)=0~,
\\[6pt]
\ih\,\fnb{\Cs\tn\k\,,\,{*}(\g\,\psi)}-{*}(\ih\,\nasl\psi-m\psi)=0~.
\end{cases}$$
By some elaboration, these can be set in the usual form of the Dirac equations,
namely
$$\begin{cases}
-\iO\,g^{ab}\,\nabla\!_a\bar\psi_{\b i}\,\g\iIi b\b\a-m\,\bar\psi_{\a i}
-\ih\,g^{ab}\,\tau_a\,\bar\psi_{\b i}\,\g\iIi b\b\a=0~,
\\[8pt]\phantom{-}
\iO\,g^{ab}\,\g\iIi a\a\b\,\nabla\!_b\psi^{\b i}-m\,\psi^{\a i}
+\ih\,g^{ab}\,\tau_a\,\g\iIi b\a\b\,\psi^{\b i}=0~,
\end{cases}$$
where $\tau_a$ is the torsion 1-form (\sref{ss:Generalized replacement principle}).

As for the gauge field, we get the field equation
$$0= -\oh\,\dO_\k{*}\dO_\k\k+\iO\,\langle{*}\g\,,\,\bar\psi\tn\psi\rangle~,$$
with the coordinate expression
$$\tfrac1\rrdg\,\de_a\bigl(\rho\Ii{abi}j\,\rdg\bigr)
+\rho\Ii{abi}h\,\k\iIi ahj-\k\iIi aih\,\rho\Ii{abh}j
-\iO\,g^{ab}\,\bar\psi_{\b j}\,\g\iIi a\b\a\,\psi^{\a i}=0~.$$

\subsubsection*{Canonical energy-tensors}

The canonical energy tensors
(\sref{ss:Currents and energy-tensors})
for the considered boson, fermion and gauge sectors have, respectively,
the expressions
\begin{align*}
&(\Ucal\spec{boson})\Ii ab=
\lambda\spec{boson}\,\d\Ii ab
-\Pi\Ii a{i\sA}\,\nabla\!_b\phi^{i\sA}-\Pi^{ai\sA}\,\nabla\!_b\bar\phi_{i\sA}=
\\[6pt]
&\qquad=\bigl(
\oh\,\bigl(g^{cd}\,\nabla\!_c\bar\phi_{i\sA}\,\nabla\!_d\phi^{i\sA})\,\d\Ii ab
-m^2\,\bar\phi_{i\sA}\,\phi^{i\sA}\,\d\Ii ab
-g^{ac}\,(\nabla\!_c\bar\phi_{i\sA}\,\nabla\!_b\phi^{i\sA}
+\nabla\!_b\bar\phi_{i\sA}\,\nabla\!_c\phi^{i\sA}) \bigr)\,\rdg~,
\\[12pt]
&(\Ucal\spec{Dirac})\Ii ab=
\lambda\spec{Dirac}\,\d\Ii ab
-\Pi\Ii a{\a i}\,\nabla\!_b\psi^{\a i}_b-\Pi^{a\a i}\,\nabla\!_b\bar\psi_{\a i}=
\\[6pt]
&\qquad=\ih\,(\bar\psi_{\a i}\,\g\iIi c\a\b\,\nabla\!_d\psi^{\b i}
-\nabla\!_c\bar\psi_{\a i}\,\g\iIi d\a\b\,\psi^{\b i})\,
(g^{cd}\,\d\Ii ab-g^{ca}\,\d\Ii db)\,\rdg
-m\,\bar\psi_{\a i}\,\psi^{\a i}\,\d\Ii ab\,\rdg~,
\\[12pt]
&(\Ucal\spec{gauge})\Ii ab=
\lambda\spec{gauge}\,\d\Ii ab+2\,\Pi\IiI{ac}ij\,\rho\iIi{bc}ij=
\bigl(\oq\,\rho\Ii{cd\,i}j\,\rho\iIi{cd}ji\,\d\Ii ab
-\rho\Ii{ac\,i}j\,\rho\iIi{bc}ji\bigr)\,\rdg~.
\end{align*}

Then it is not difficult to check that in all cases the canonical energy-tensor
and the stress-energy tensor are related by
$$\Ucal_{ab}+\Ucal_{ba}=-4\,\Tcal_{ab}$$
(the symmetrization in the indices $a$ and $b$ is only required
for the Dirac field),
and that by evaluation through critical fields one gets
$$\codiv(\Tcal\!\!\spec{matter}+\Tcal\!\!\spec{gauge})=\text{torsion terms}.$$
We also note that $\Tcal\!\!\spec{matter}$ and $\Tcal\!\!\spec{gauge}$
are not separately divergence-free: their sum is such.

\subsubsection*{Gravitational field}
The ``metric-affine'' approach to gravity can be treated
in the covariant-differential formalism, too.
Let the gravitational field be represented by the couple $\bigl(g,\G\bigr)$
constituted by a spacetime metric and a linear spacetime connection,
and \hbox{$\Lambda\spec{grav}=\lambda\spec{grav}\,\dO^4\xx$} with
$$\lambda\spec{grav}=g^{ad}\,\d\Ii bc\,R\iIi{ab}cd\,\rdg~,\qquad
R\equiv-\dO_\sGa\G~.$$
Then we find \hbox{$\Pi^{\sst(0)}=G\tn\eta$}\,,
where $G$ is the Einstein tensor, \hbox{$\Pi^{\sst(1)}=0$}\,,
and
$$\Pi\IiI{ab}cd\equiv(\Pi^{\sst(2)})\IiI{ab}cd=
\oh\bigl(g^{bd}\,\d\Ii ac-g^{ad}\,\d\Ii bc\bigr)\,\rdg~.$$
Accordingly, the field equations (theorem~\ref{theorem:fieldequations})
turn out to be the Einstein equation for the $g$-sector and the equation
\begin{align*}
&\dO_\G\bigl(\Pi^{\sst(2)}\comp(g,\G)\bigr)=0
\\[6pt]\text{i.e.}\quad
&\de_a\Pi\IiI{ab}cd-\G\iIi ade\,\Pi\IiI{ab}ce+\G\iIi aec\,\Pi\IiI{ab}ed=0~,
\end{align*}
for the $\G$-sector.
After some elaboration this can be written in the form
$$\nabla\!_c(g^{bd}\,\rdg)=\text{torsion terms}.$$
If the torsion is assumed to vanish then this is equivalent
to \hbox{$\nabla\!_cg^{bd}=0$}\,.



\begin{thebibliography}{10}
%
\bibitem{C07}
D.\ Canarutto:
`{}``Minimal geometric data'' approach to
Dirac algebra, spinor groups and field theories',
Int.\ J.\ Geom.\ Met.\ Mod.\ Phys., {\bf 4} N.6, (2007), 1005--1040.\\
arXiv:math-ph/0703003.
%
\bibitem{C16c}
D.\ Canarutto:
`Overconnections and the energy tensors of gauge and gravitational fields',
J.\ Geom.\ Phys.\ {\bf 106} (2016), 192--204.\\
doi:10.1016/j.geomphys.2016.03.027 
%
\bibitem{CM85}
D.\ Canarutto and M.\ Modugno:
`Ehresmann's connections and the geometry of energy-tensors
in Lagrangian field theories',
Tensor {\bf 42} (1985), 112--120.
%
\bibitem{Costantini94}
P.G.\ Costantini:
`On the geometrical structure of Euler-Lagrange equations',
Annali di Matematica Pura e Applicata (IV), Vol.\ CLXVII (1994), 389--402.
%
\bibitem{FernandezGarciaRodrigo00}
A.\ Fern\'andez, P.L.\ Garc{\'\i}a and C. Rodrigo:
`Stress-energy-momentum tensors in higher order variational calculus',
J.\ Geom.\ Phys.\ {\bf 34} N.1 (2000), 41--72.
%
\bibitem{FerrarisFrancaviglia92}
M.\ Ferraris and M.\ Francaviglia:
`Conservation laws in general relativity',
Class. and Quantum Grav. {\bf 9} (1992), S79.
%
\bibitem{FatibeneFerrarisFrancaviglia94}
L.\ Fatibene, M.\ Ferraris and M.\ Francaviglia:
`Noether formalism for conserved quantities in classical gauge field theories',
J.\ Math.\ Phys.\ {\bf 35} 1644 (1994).
%
\bibitem{ForgerRoemer04}
M.\ Forger and H.\ R\"omer:
`Currents and the energy-momentum tensor in classical field theory:
a fresh look at an old problem',
Annals Phys.\ {\bf 309} (2004), 306--389;
arXiv:hep-th/0307199.
%
\bibitem{FroNij56}
A.\ Fr\"olicher and A.\ Nijenhuis: 
\emph{Theory of  vector--valued differential forms, Part I}, 
Indagationes Mathematicae {\bf 18} (1956), 338--360.
%
\bibitem{FroNij60}
A.\ Fr\"olicher and A.\ Nijenhuis:
\emph{Invariance of vector form operations under mappings},
Commentarii Mathematici Helvetici {\bf 34} (1960), 227Ð-248.
%
\bibitem{Gar74}
P.L.\ Garc{\'\i}a:
`The Poincar{\'e}-Cartan Invariant in the Calculus of Variations',
Symposia Mathematica {\bf 14} (1974), 219--246.
%
\bibitem{GS73}
H.\ Goldsmith and S. Sternberg:
`The Hamilton-Cartan formalism in the calculus of variations',
Ann.\ Inst.\ Fourier, Grenoble {\bf 23} (1973), 203--267.
%
\bibitem{GotayMarsden92}
M.J.\ Gotay and J.E.\ Marsden:
`Stress-energy-momentum tensors and the Belinfante-Rosenfeld formula',
Contemp.\ Math. {\bf 132} (1992), 367--392.
%
\bibitem{Grabowski13}
J.\ Grabowski: `Brackets',
Int.\ J.\ Geom.\ Methods Mod.\ Phys.\ {\bf 10}, 1360001 (2013) [45 pages],
DOI: http://dx.doi.org/10.1142/S0219887813600013\\
arXiv: 1301.0227 [math.DG]
%
\bibitem{HE73}
S.W.\ Hawking and G.F.R.\ Ellis:
\emph{The large scale structure of space-time},
Cambridge Univ.\ Press, Cambridge (1973).
%
\bibitem{Hermann75}
R.\ Hermann:
`Gauge fields and Cartan-Ehresmann connections, Part A',
\emph{Interdisciplinary Mathematics} {\bf X},
Math.\ Sci.\ Press, Brooklyn (1975).
%
\bibitem{HorakKolar83}
M.\ Hor\'ak and I.\ Kol\'a\v{r}: `On the higher order Poincar\'e-Cartan forms',
Czechoslovak Math.\ J.\ {\bf 33} (1983), 467Ð-475.
%
\bibitem{Janyska07}
J.\ Jany\v{s}ka:
`Higher-order Utiyama invariant interaction',
Rep.\ Math.\ Phys.\ {\bf59} (2007), N.1, 63--81.
%
\bibitem{Kolar84}
I.\ Kol\'a\v{r}:
`A geometrical version of the higher order Hamilton formalism in fibred manifolds',
J.\ Geom.\ Phys.\ {\bf 1}, n. 2 (1984), 127--137.
%
\bibitem{KolarMichorSlovak93}
I.\ Kol\'a\v{r}, P.\ Michor, and J.\ Slov\'ak:
\emph{Natural Operations in Differential Geometry},
Springer-Verlag (1993).
%
\bibitem{Kos50}
J.\ L.\ Koszul: 
`Homologie et cohomologie des alg\`ebres de Lie', 
Bulletin de la S.M.F.\ {\bf 78} (1950), 65--127.
%
\bibitem{Kos60}
J.\ L.\ Koszul: 
\emph{Lectures on fibre bundles and differential geometry}, 
Tata Institute, Bombay (1960).
%
\bibitem{Krupka02}
D.\ Krupka: `Variational principles for energy-momentum tensors',
Rep.\ Math.\ Phys.\ {\bf49} (2002), N.2--3, 259--268.
%
\bibitem{Krupka15}
D.\ Krupka: \emph{Introduction to global variational geometry},
Springer (2015).
%
\bibitem{KrupkaKrupkovaSanders10}
D.\ Krupka, O.\ Krupkov\'a and D.\ Saunders:
`The Cartan form and its generalizations in the calculus of variations',
Int.\ J.\ Geom.\ Met.\ Mod.\ Phys.\ {\bf 7} N.4, (2010), 631--654.
%
\bibitem{Kuperschmidt79}
B.\ Kuperschmidt, `Geometry of jet bundles and the structure
of Lagrangian and Hamiltonian formalisms',
Lecture Notes in Math., {\bf 775} (1979), 162--218.
%
\bibitem{LandauLifchitz68}
L.\ Landau, E.\ Lifchitz:
\emph{Th\'eorie du champ}, Editions Mir, Moscou (1968).
%
\bibitem{Leclerc06}
M.\ Leclerc:
`Canonical and gravitational stress-energy tensors',
Int.\ J.\ Mod.\ Phys.\ D15 (2006) 959-990;
arXiv:gr-qc/0510044.
%
\bibitem{MaMo83b}
L.\ Mangiarotti and M.\ Modugno:
`Some results on the calculus of variations on jet spaces',
Ann.\ Inst.\ H.\ Poinc.\ {\bf 39} (1983), 29--43.
%
\bibitem{ManMod84}
L.\ Mangiarotti and M.\ Modugno:
\emph{Graded Lie algebras and connections on a fibred space}, 
Journ.\ Math.\ Pur.\ et Applic.\ {\bf 63}, (1984), 111--120.
%
\bibitem{Michor01}
P.\ W.\ Michor: `Fr\"olicher-Nijenhuis bracket', in M.\ Hazewinkel,
\emph{Encyclopedia of Mathematics}, Springer (2001), ISBN 978-1-55608-010-4.
%
\bibitem{Mod91}
M.\ Modugno: 
\emph{Torsion and Ricci tensor for non linear connections},
Diff.\ Geom.\ Appl.\ {\bf 1} (1991), 177--192.
%
\bibitem{Nij72}
A.\ Nijenhuis:
`Natural bundles and their general properties',  
in Diff.\ Geom., in honour of K. Yano, Kinokuniya, Tokyo (1972), 317--334.
%
\bibitem{ObukhovPuetzfeld14}
Y.N.\ Obukhov and D.\ Puetzfeld:
`Conservation laws in gravity: A unified framework',
Phys.\ Rev.\ D {\bf 90} (2014), 024004;
arXiv:1405.4003 [gr-qc].
%
\bibitem{Pons09}
J.M.\ Pons: `Noether symmetries, energy-momentum tensors and conformal invariance
in classical field theory',
J.\ Math.\ Phys.\ {\bf 52}, 012904 (2011);\\
http://dx.doi.org/10.1063/1.3532941.
%
\bibitem{Sa89}
 D.\ J.\ Saunders:
\emph{The Geometry of Jet Bundles},
Cambridge University Press (1989).
%
\bibitem{Tr67}
A.\ Trautman:
`Noether Equations and Conservation Laws',
Commun.\ Math.\ Phys.\ {\bf 6} (1967), 248--261.
%
\bibitem{Utiyama56}
R.\ Utiyama: `Invariant theoretical interpretation of interaction',
Phys.\ Rev.\ {\bf 101} (1956), 1597--1607.
%
\bibitem{VinogradovCSS84I_II}
A.\ M.\ Vinogradov:
`The C-spectral sequence, Lagrangian formalism, and conservation laws.
I. The linear theory',
J.\ Math.\ Anal.\ Appl. {\bf100} N.1 (1984), 1--129.
%
\bibitem{Vitolo91}
R.\ Vitolo:
`A new infinite order formulation of variational sequences',
Arch.\ Math.\ Un.\ Brunensis {\bf 34}~N.4 (1998), 483--504.
%
\bibitem{Yano55}
K.\ Yano:
\emph{Lie derivatives and its applications},
North-Holland, Amsterdam (1955).

\end{thebibliography}
\end{document}